**Interaction between Rydberg Excitons in Cuprous Oxide**
**Revealed through Resonant Second Harmonic Generation**

Andreas Farenbruch[1], Henje Stolz[2], Peter Grünwald[2], Dirk Semkat[1], Nikita Siverin[1], Dmitri R. Yakovlev[1], Dietmar Fröhlich[1], and Manfred Bayer[1]

[1]Experimentelle Physik 2, Technische Universität Dortmund, 44221 Dortmund, Germany,
[2]Institut für Physik, Universität Rostock, Albert-Einstein-Str. 23, 18059 Rostock, Germany,

**Abstract**

We report experimental and theoretical investigations of interacting excitons of the yellow series in cuprous oxide ($Cu_2O$) with principal quantum numbers up to $n=7$ by means of second harmonic generation (SHG). Using picosecond pulsed laser excitation up to 10 GW/cm$^2$ peak intensity we observe a pronounced change of the spectra with increasing pump laser intensity: an energetic shift to lower absolute energies and a spectral broadening. The absolute intensities of the spectral lines scale for low powers with the square of the pump power, but saturates at higher powers. At still higher powers the SHG intensity is actually reduced. To explain these results quantitively, we developed a semi-classical theory of resonant SHG where the process of SHG is fully coherent. The excitons are assumed to be bosons interacting by a distance dependent potential $V(r)$ giving rise to both the changes in spectral line shape and the saturation by a 'Rydberg' blockade. The concomitant measurement of two-photon absorption allows to derive quantitative values for the exciton-exciton interaction. While the results agree in order of magnitude with those calculated by state-of-the art atomic-like van der Waals interaction theory, the scaling with principle quantum number is quite different. As a possible screening by an electron-hole plasma created by three-photon absorption into blue and violet band states could be ruled out, our results point toward fundamental differences between excitons and atoms.

## I. Introduction

Since their first observation, Rydberg excitons in cuprous oxide ($Cu_2O$) [1] have turned out to be fascinating quantum objects with interesting properties ranging from quantum chaos [2] to strong Kerr nonlinearities [3] (for reviews see [4,5]). Their similarity to Rydberg atoms [5] implies a strong dipole-dipole interaction leading to the phenomenon of Rydberg blockade [7]. Indeed, the strong changes in the one-photon absorption (1PA) spectrum with respect to the exciting laser power as observed in [1], were explained by this blockade effect. Recently, experiments demonstrated the *asymmetric blockade*, i.e., the change in the spectrum of one Rydberg state due to the pumping of another different Rydberg state. The concomitant theoretical analysis showed that these results are consistent with an atomic-like van der Waals interaction law for the excitons [8].

In all of these studies, the Rydberg excitons have been created by one-photon absorption in thin single crystalline platelets of $Cu_2O$ using continuous-wave (cw) excitation, preferentially by a single-frequency laser [1,9,10]. Due to the equal parity of the electronic bands from which the excitons originate, which are the $\Gamma_7^+$ valence and the $\Gamma_6^+$ conduction band [4], the strongest transitions belong to excitons with an $L=1$ envelope (P states) [11]. The interaction of the yellow 1S excitons with odd parity optical phonons results in a strong phonon sideband [12], which superimposes the P absorption lines and leads to their well-known asymmetry [13]. While this peculiarity does not affect experiments where by pump-probe schemes the interactions between Rydberg excitons have been investigated [1,8], another property of $Cu_2O$ has turned out to be quite problematic in these studies, the process of Auger-like scattering between two excitons [14,15]. In this process two yellow excitons, which can be both in different spin states, named conventionally ortho and para states [16], scatter at each other, whereby the energy of one of the excitons is transferred to the other resulting in an unbound electron-hole pair. On the one hand, this Auger process limits the maximum possible exciton density [16], on the other hand, it leads to the creation of an electron-hole plasma (EHP), even at very low excitation laser powers, whenever yellow 1S excitons are created via the phonon assisted absorption. This inevitably is the case in every scheme involving one-photon absorption [16]. In a typical experiment [1,8,18] application of a laser power of 100 μW would lead to EHP densities of the order of $10^{10}\ \mathrm{cm}^{-3}$ and plasma temperatures around 10 K [18,19]. The standard description for a low-density plasma, the Debye model [20], predicts that plasmas with such densities should lead to a negligible effect on Rydberg exciton states at principal quantum numbers below $n=20$ [18,21]. However, as shown recently, the Debye model is not applicable to such low temperature EHP [22,23] and even at very low concentrations an EHP leads to substantial effects on the absorption line shape of the P excitons and may mask completely the expected Rydberg blockade of Rydberg excitons [22,23,19]. So, in order to be able to unambiguously demonstrate the interaction between Rydberg excitons one should look for scenarios in which the creation of an EHP made up from the yellow band states ("yellow" plasma) is suppressed. The rate models developed in [16,19] imply a significant delay for the creation of Auger-based EHP. Hence, one approach would be to switch from cw-excitation to subnanosecond pulses and to observe the changes in the absorption spectra of the P excitons with time. Another obvious way to avoid direct excitation of yellow 1S excitons is using the well-known two-photon excitation into S and D states with higher principal quantum number [24,25], as here any process involving odd parity optical phonons is symmetry forbidden. In such a scenario 1S excitons and thus Auger decay would come up only via the decay of the Rydberg states, i.e., after times of the order of their lifetimes (see Fig. 1), so an influence of a yellow EHP can be diminished.

Utilizing as detection channel for the excitons the quadrupole emission one arrives at a process called "resonant second harmonic generation" (RSHG) [26]. Indeed, using femtosecond laser pulses it was shown that in this way exciton states with principal quantum numbers up to $n=9$ could be detected [26-28]. However, these studies have been focused on the physical mechanisms by which these processes become observable, i.e., the symmetry and polarization properties and on the influence of a magnetic field. Information on density dependent effects, i.e., on exciton-exciton interaction has not been obtained. These experiments also face the problem that due to the large spectral bandwidth of the femtosecond laser pulses many Rydberg states are excited simultaneously leading to interference effects [27,28]. Reducing the spectral bandwidth by employing nanosecond laser pulses allowed to observe well-resolved Rydberg states in SHG with principal quantum numbers up to $n=12$ [29]. In these studies, no effects of an exciton-exciton interaction could be detected, because due to the rather long pulse duration the maximum laser power that could be applied before thermal heating of the sample sets in was far too low.

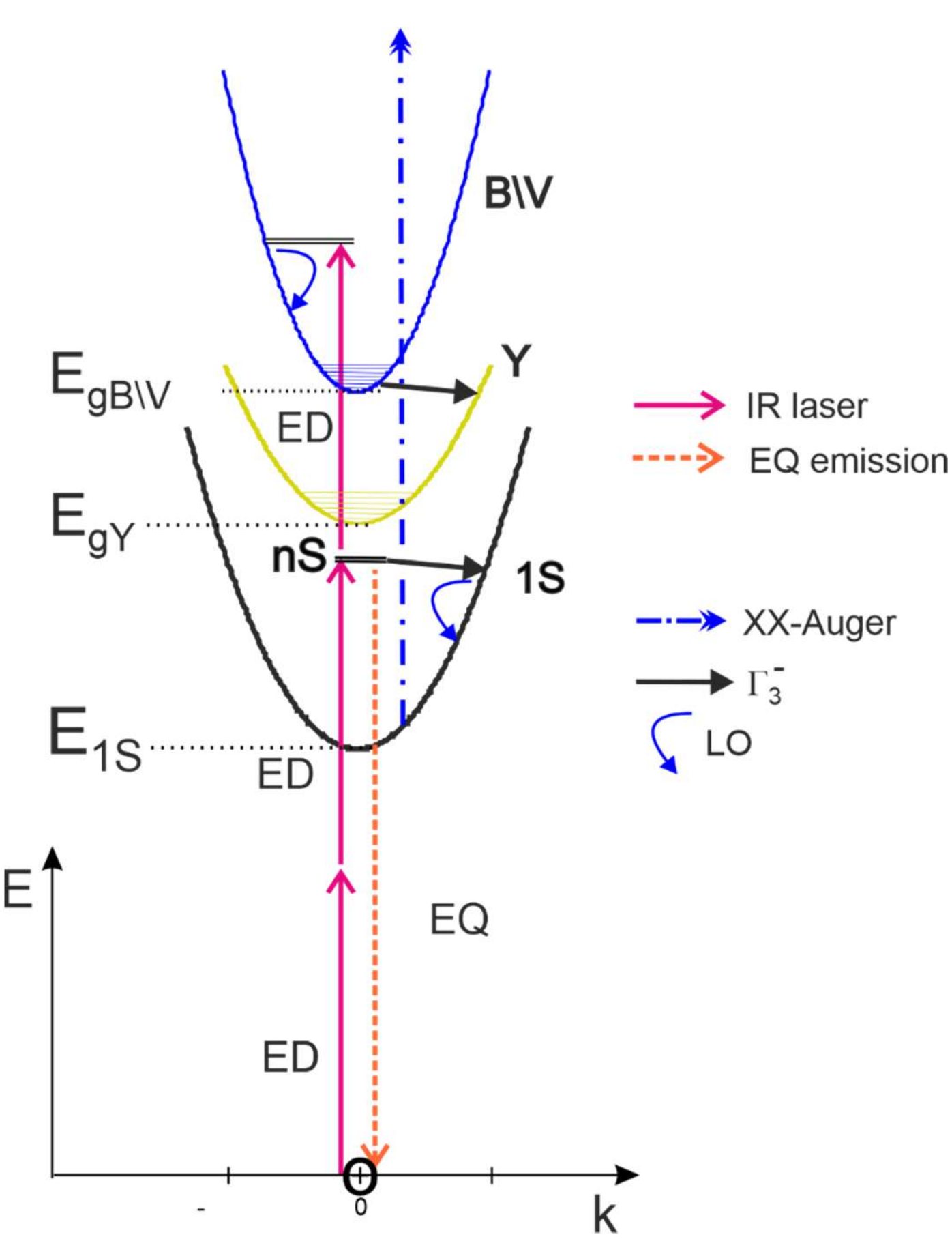

Fig. 1: Energy scheme of the important optical and relaxation processes in $Cu_2O$ in a two-particle picture.
O designates the crystal ground state, the parabolas the energy dispersion of the yellow 1S exciton (1S, black), the yellow band continuum (Y, yellow) and the blue/violet continuum (B\V, blue), the corresponding band gap energy, i.e. $E_{gY}$ , $E_{gB\V}$ , is indicated by a dotted black line.
The black double lines denote the final states of (i) the two-photon (nS exciton) and of (ii) the three-photon absorption process. ED, EQ denote an electric dipole or quadrupole transition, respectively. The thick black arrows denote scattering by optical $\Gamma_3^-$ phonons, the light blue arrows by LO phonons. The blue dashed-dotted arrow indicates the bimolecular Auger-like process resulting in electron-hole pairs in high yellow and blue continuum states. The hatched areas near the bottom of the band continua denote electron-hole plasmas.

In all these studies, however, the question of what happens if one excites the crystal outside any resonance, i.e., the possible existence of a nonresonant SHG process described by a susceptibility $\chi^{(2)}$, was left open. That such a process must be possible is shown by the straightforward argument, that the inclusion of higher-order exciton-photon interactions of the valence and conduction bands of appropriate symmetries allows SHG despite the centrosymmetry of $Cu_2O$, see e.g., the study of SHG of blue and violet exciton states [30]. This of course means that the spectral dependence of $\chi^{(2)}$ is independent of the pump field intensity, leading to a strict quadratic dependence on pump power and it must be almost flat giving rise to a spectrum that is identical with the absolute square of the self-convolution of the electric field of the pump pulse [31]. This can be experimentally determined simply by measuring the SHG of the pump pulse using a broadband frequency doubler like BBO.

Furthermore, RSHG was considered to be a *coherent* process, i.e., that the electric field of the SHG light is directly related to the amplitude of the resonant exciton states as obviously shown by the quantum optical theory of Ref. 29. Here, we expect as a fingerprint that the SHG process will be influenced by propagation and stimulated emission effects [32].

A recent time-resolved study of SHG in $Cu_2O$ [33] showed that for low $n$ Rydberg states of S type a close agreement between lifetime $T_1$ and spectral line width $\hbar\Gamma=\hbar/T_1$ could be found, but not for D states. These studies put forward the question whether the previous SHG experiments really showed the coherence of the process. Indeed, RSHG, like any other resonant process, e.g., resonant Raman scattering [34], can be considered as a form of secondary emission (SE) and described as an *absorption followed by emission* (AfE) process. Hereby, in a first step excitons are created by a two-photon excitation and then emit light by the quadrupole transition [25]. Obviously, such a description includes both coherent and incoherent contributions, but addresses the question of coherence by, e.g., quantum beats [34].

The relation of both descriptions, i.e., the details of coherent and incoherent contributions is, however, a quite complex problem (see e.g. Ref. 35 to 37) and depends not only on the relaxation processes of the intermediate states [38], but also on the methods of excitation and detection of the SHG signal [34]. In essence, however, it turns out that the simple model of AfE is only valid in the limiting case of broadband excitation and detection and under stationary conditions [34,36]. In the latter case the spectrally integrated intensity $S_X(P_L,n)$ is proportional to the two-photon absorption constant, the square of pump power ($\chi^{(2)}$ process) and the probability of the radiative process $w_{\mathrm{rad}} = \Gamma_{\mathrm{rad}} / \Gamma_{\mathrm{tot}}$

$$S_X(P_L,n) \propto \rho_X \cdot w_{\mathrm{rad}} \propto \alpha_{TPA}(n) P_L^2 \frac{\Gamma_{\mathrm{rad}}(n)}{\Gamma_{\mathrm{tot}}(n)}, \tag{1.1}$$

while the spectrum is just a Lorentzian of a width given by the dephasing time of the emitting state.

In this paper, we apply SHG using picosecond laser pulses with intensities up to the GW/cm$^2$ range tuned into resonance, i.e., twice the central pulse laser frequency, with the yellow exciton states expecting large enough exciton densities to study the interaction between excitons. The SHG light was analyzed using conventional spectroscopy giving the time-integrated spectrum, which allows a well-defined theoretical analysis. Indeed, in our experiments we were able to observe RSHG by exciton states with principal quantum numbers up to $n = 7$ as in previous studies [27,28] and could discriminate it from a nonresonant contribution. The resonant spectra show large non-linear effects like changes in the line shape and an intensity saturation. To understand the spectral signatures, we developed a theory of the resonant SHG of excitons that takes into account: i) the bosonic character of excitons, ii) the propagation of both the pump and the second harmonic light in the sample including their wave vector mismatch, iii) exciton relaxation processes, and iv) the exciton-exciton interaction.

To fully understand the experimental results, we also have to consider that the excitons are in a solid-state environment, interacting with many different quasi-particles, e.g., phonons and various electron-hole plasmas of the yellow and blue/violet conduction bands schematically depicted in Fig. 1. The plasmas may be optically generated by three-photon absorption (3PA) [39] into the blue and violet continuum producing on the time scale of the pump pulses free electrons and holes in these bands (called “blue-violet” EHP, see Fig. 1). This will directly interact with the yellow Rydberg excitons leading to additional spectral broadening due to phonon-plasmon scattering [19,40]. As this process will increase with the third power of the pump intensity, we can separate the effect of such a blue-violet EHP on the Rydberg exciton states from the direct exciton-exciton interaction, which should depend on the exciton density, i.e., on the second power of the pump intensity.

However, the prerequisite for a quantitative determination of the exciton-exciton interaction strength that would allow for a stringent test of any theory is the independent determination of the exciton density realized in the experiment. Here, the method of SHG is also unique, as the number of absorbed photons of the infrared (IR) laser, given by the loss of the laser energy in the sample, is exactly half of the number of excitons during the pulse, thus allowing to derive the two-photon absorption (2PA) coefficient. The results obtained should allow a quantitative comparison.

The paper is organized as follows. In Section II, we describe shortly the experimental setup and exemplarily present the results for SHG measurements. We will also demonstrate the decomposition of the data into resonant and nonresonant contributions, which turns out to be essential for the analysis. In the next section we present the theoretical model to analyze the experiment. Section IV presents the analysis of the spectra by fitting the theoretical model to the line shapes and deriving in this way the dependence of total intensity, line shift and line width on pump power. This allows to derive quantitative values for the exciton-exciton interaction scaling laws and to discuss the results within a Rydberg blockade model. In two appendices we discuss details of the 2PA measurements and the reconstruction of the temporal electric field of the pump pulse. The paper closes with conclusions and an outlook.

## II. Experimental Methods and Results

In our experiments SHG is measured by using for excitation a tunable ps-pulse laser source with a spectral width (FWHM) of about 0.9 meV. The repetition rate was set to 30 kHz and the average laser power $P_L$ could be varied from 1 mW to 100 mW. The Gaussian laser beam is focused to about 120 µm beam waist on the crystal sample immersed in superfluid helium at $T = 1.4$ K. The thickness of the well-polished sample was determined by interferometry to be $(29 \pm 1)$ µm, but shows a slight wedge of about 0.3°. The excitation and detection polarization are chosen in such a way that the SHG is allowed for the even parity states (for details see Refs. 27,28). In Figure 2, panel a) the spectrum of the second harmonic of the pump pulse is shown, which is representative for the nonresonant SHG to be expected (see below). Obviously, its line shape is rather complex but actually this helps to identify the nonresonant contributions. Since the SHG is critically dependent on the temporal phase of the pump pulse, we reconstructed the temporal field strength from the measured spectrum as detailed in Appendix A. It turns out that the time-dependence of the pump pulse intensity can be well described by a square pulse with pulse duration of 3.6 ps with an almost flat temporal phase. This temporal profile was then used in the theoretical calculations.

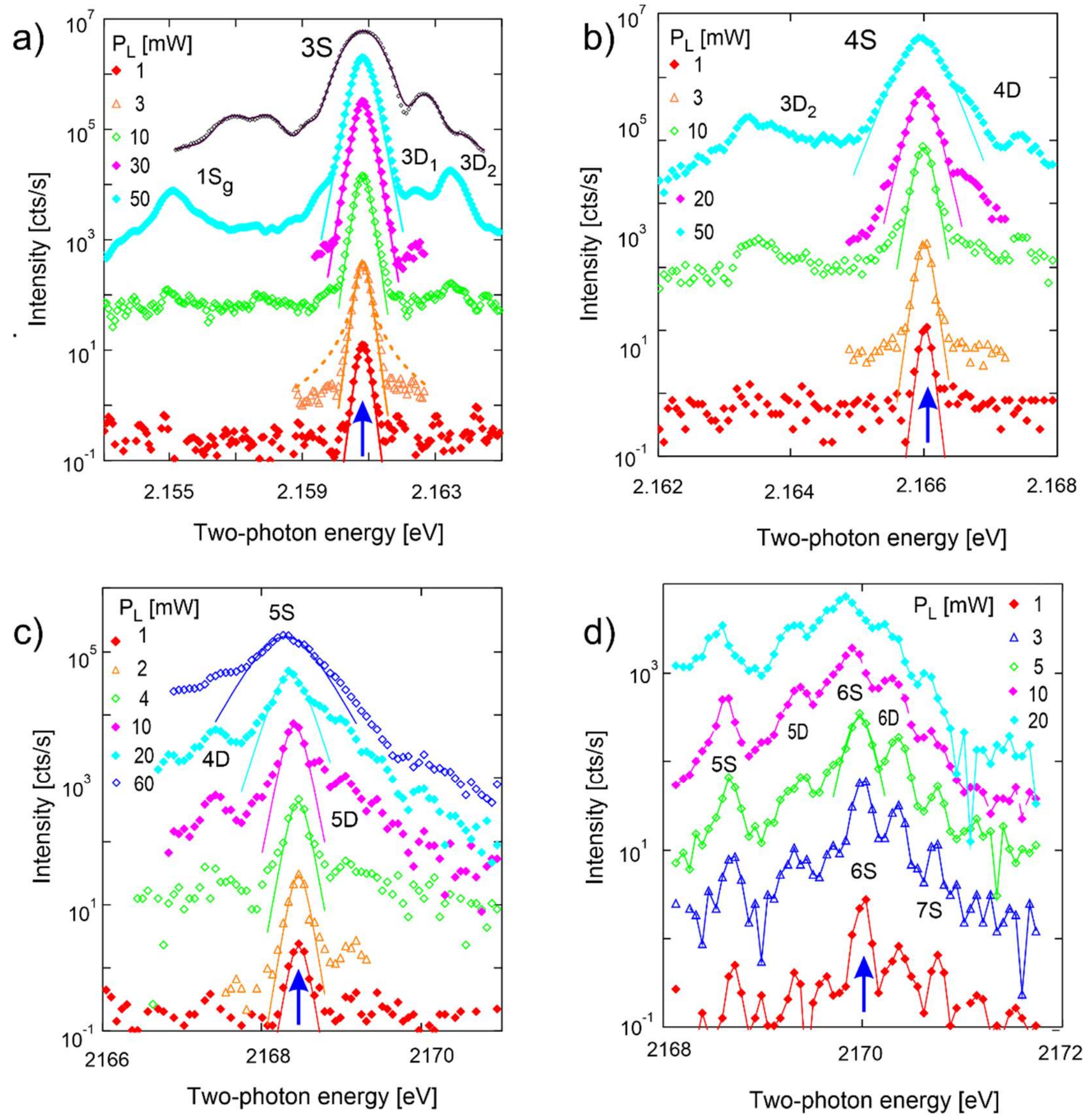

Fig. 2: SHG spectra (i.e., emitted light intensity vs. photon energy after subtraction of the dark noise of the camera) for excitation around the positions indicated by the blue arrows (for a) $2E_L = 2.1609\,\mathrm{eV}$, b) $2E_L = 2.1660\,\mathrm{eV}$, c) $2E_L = 2.1685\,\mathrm{eV}$, and d) $2E_L = 2.1700\,\mathrm{eV}$). The average pump powers are varying between 1 mW and 60 mW as indicated. To enhance the visibility the spectra are scaled with a factor $\sqrt{10}$ with increasing pump power. In contrast to the expectations, the spectral lineshapes are not given by a Lorentzian (as demonstrated by the ochre dashed line in part a) for a power of 3 mW) but by a "Secans hyperbolicus (SECH)" shape as given by the full lines in the spectra, which are fits with this lineshape. Besides the dominant nS line (n ranging from 3 in panel a) to 6 in panel d)), at higher pump powers we can identify weak transitions involving other exciton states (in a) the 1S green exciton and two 3D yellow states, in b) and c) 3D and 4D states, while in d) all exciton states between 5S and 7S show up. In panel a) the black diamonds give the spectrum of the second harmonic of the pump laser, the black full line is a fit with a sum of 10 gaussians.

We then measured the spectra (spectral resolution $60\,\mu\text{eV}$) of the light emitted from the sample, spatially filtered so that only the light within the exciting laser beam is collected. In all spectra a background due to the dark response of the photodetector is subtracted.

By tuning the excitation laser pulse in resonance with the various exciton transitions we measured the SHG spectra for S lines for the $n=3$ to $n=7$ excitons as a function of laser power. Some selected results are displayed in Fig. 2, where the laser pulse was centered at the positions indicated by blue arrows. We see that all resonances show a common behavior: at low excitation power ($P_L$ = 1 mW) the lines have a rather small linewidth (red curves). With increasing pump power, the lines broaden and shift to lower energies, e.g., compare the 5S line at 1 mW and 20 mW. The magnitude of the shift strongly depends on the principal quantum number $n$, becoming quite large for the highest $n$. Most remarkable, however, is the *line shape*, which at low powers can be fitted *only* with a secans hyperbolicus function $S(\omega) = S_0\, sech\left((\omega - \omega_0)/s\right)^2 / 2s$ (see colored solid lines in Fig. 2) and obviously not with the usual Lorentzian (see the ocher dashed line at 3 mW in Fig. 2a). While for the 3S line this shape remains up to the highest power level and stays always smaller than the width of the laser pulse, the other lines show a different behavior. Their broadening at the highest power levels becomes quite large (see the spectra at the highest power in Fig. 2c) and resembles that of the SHG of the pump laser (black line in Fig. 2a). Obviously, this gives a strong hint that at high powers the SHG process is dominated by the nonresonant contribution. Therefore, in a first step of the analysis, we identified in the spectra the nonresonant contribution. By using the facts that its spectrum is identical to the known SHG spectrum of the IR pulse and that it should be the same for all lines and scale with $P^2$, it turned out to be reliably and consistently possible. Examples are shown in Fig. 3 for different laser powers and exciton states. We see that the nonresonant contribution in the low power range is quite small and can be completely neglected for n=3 and 4. In the high-power range, the spectrum is dominated by the nonresonant contribution, but nevertheless the decomposition is reliable as can be seen by the results for nS=6 and 30 mW.

In this way one retrieves nearly background free spectra, from which the total intensity of the interesting line can be obtained by numerical integration, the results of which is shown by the points in Fig. 4 for *nS=3* to *7*. The nS=3 state is special, as here the nonresonant background contributes extremely little to the spectrum. For all exciton states we observe the expected quadratic dependence on pump power at low power range, but also a saturation behavior at

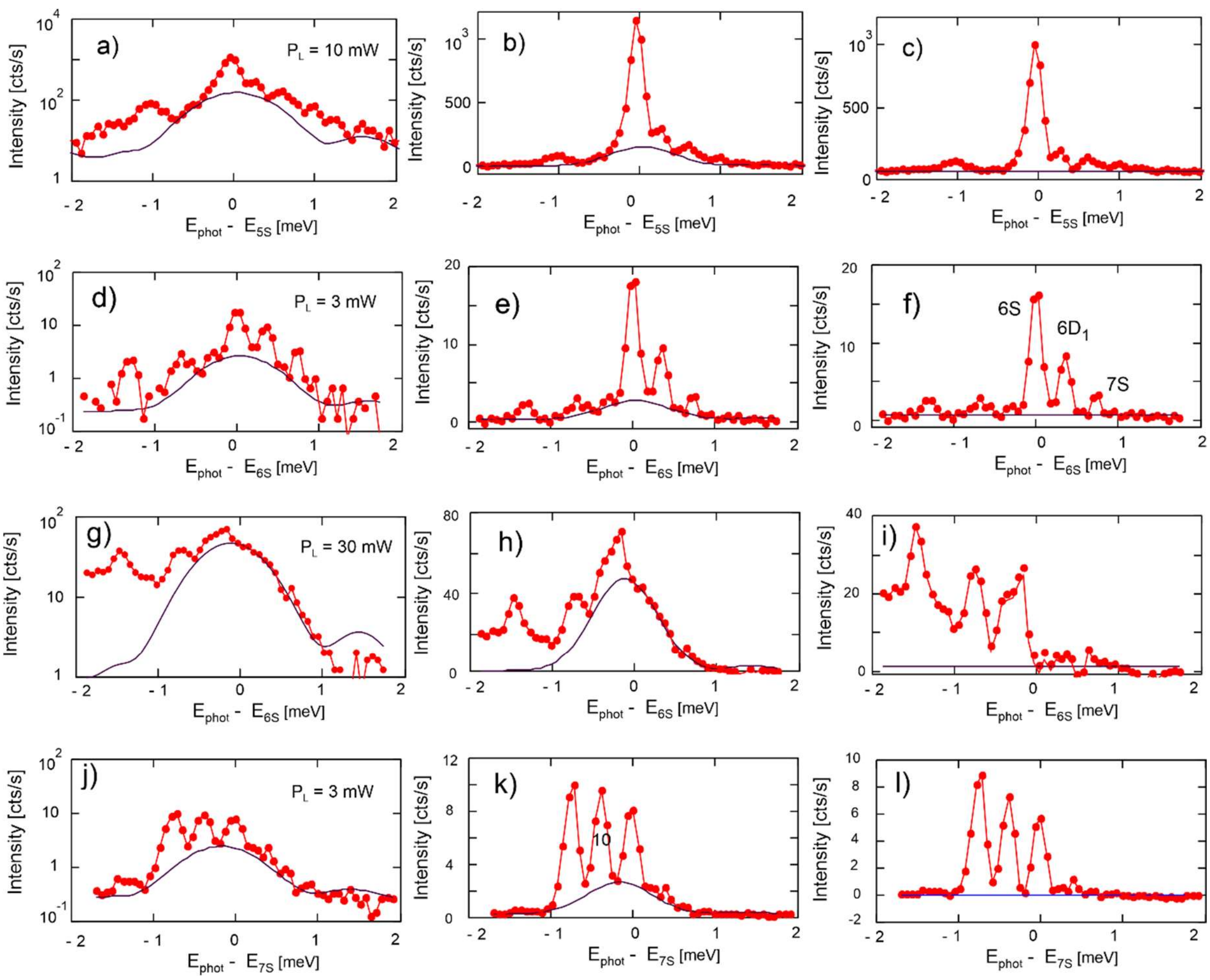

Fig. 3: Decomposition of spectra into the nonresonant and resonant contributions. The first column of panels a, d, g, and j show the experimental data (red dots) on a logarithmic and panels b, e, h, and k on a linear scale of the ordinate. The black full line gives the estimated nonresonant SHG that adheres optimal to the lower boundary of the spectrum. Panels c, f, i, and l show the resonant contribution that results from subtracting the nonresonant part from the measured spectrum. Panels a,b, c are for nS=5 and 10 mW, panels d, e, f for nS=6 and 3mW, while panels g, h, i for nS=6 and 30 mW and panels j, k, l for nS=7 and 3 mW.

high powers, which can be described by

$$S_0(n,P) = \frac{C}{n^3}\frac{P^2}{1+P^2/P_c^2} \ . \qquad (1.2)$$

For the states with *nS>4* this saturation goes over into a real loss of intensity. From Eq. (1.1) one can conclude that the only explanation would be the opening of an additional decay channel for the exciton states at high pump powers, e.g., due to phonon-plasmon scattering [19]. It is quite straightforward to assume an EHP produced by 2PA into the blue/violet bands as the origin. In any case, the density of such a plasma should increase with the third power of the pump power, so one would assume a power dependence of C of the form

$$C(P) = \frac{C_0}{1+\left(\frac{P}{P_{sc}}\right)^3} \ . \qquad (1.3)$$

Indeed, as seen by the solid lines, all data can be fitted by this relation (together with Eq. (1.2) ) very well. The parameters used for the fits are given in Table 1. Note that since the two-

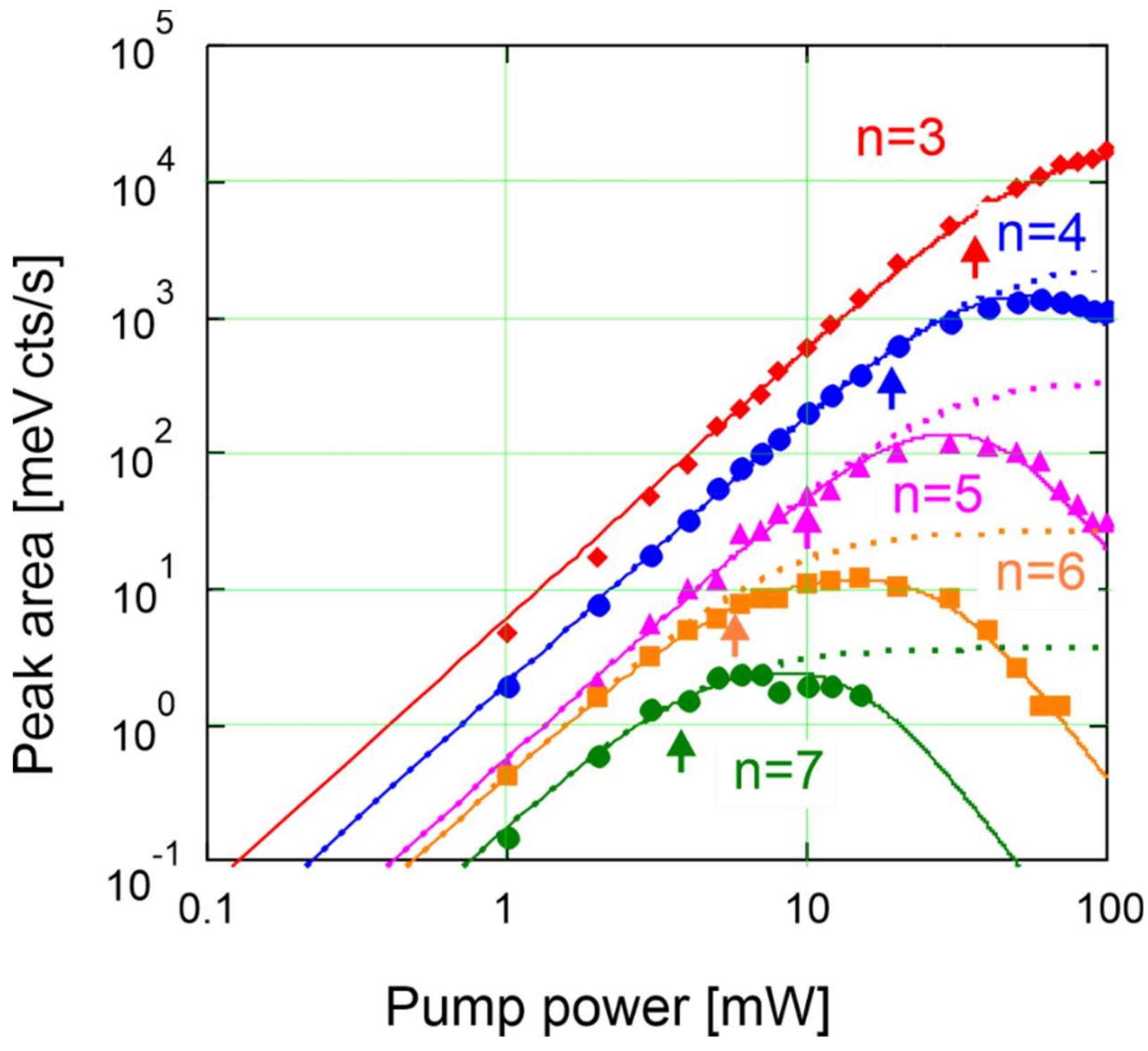

Fig. 4: Power dependence of the resonant contribution for nS=3 (red diamonds and lines), nS=4 (blue circles and lines), nS=5 (magenta triangles and lines), nS=6 (ochre squares and lines), and nS=7 (green circles and lines). The dotted lines are fits with Eq. (1.2) taking only saturation into account, while the full lines take additionally a reduction of the lifetime (see Eq. (1.3)) into account. The parameters of the fits are given in Table I. The coloured arrows denote the onset of b/v plasma induced scattering as discussed in Section V.

photon excitation as the first step of SHG, the radiative rate and the total decay rate (if

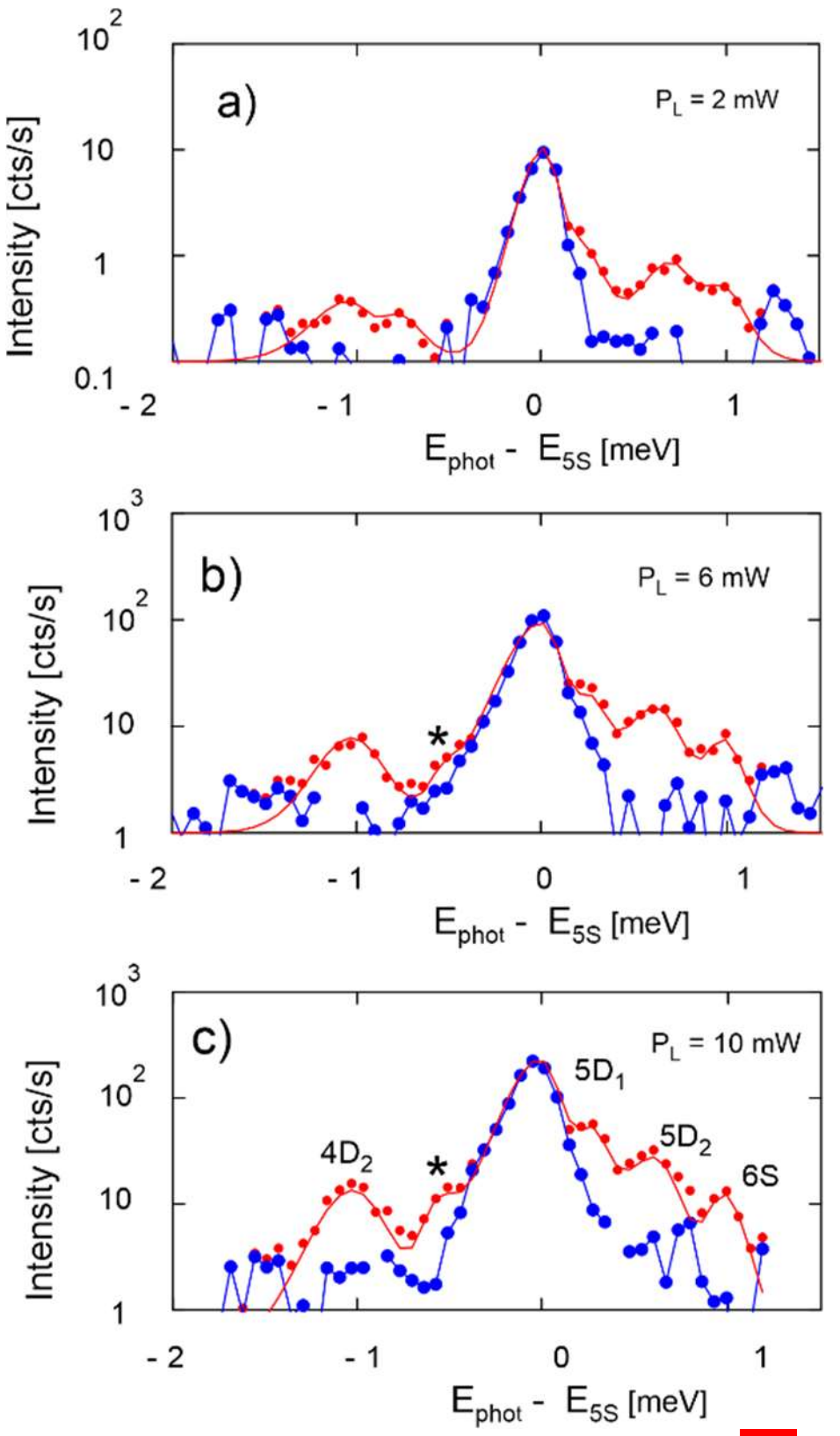

Fig. 5: Extraction of the „true" line shape of the 5S resonance for different pump powers. In each panel, the red dots give the measured spectrum, the full red line the fit and the blue dots and lines the pure spectrum of the resonant line. The additional spectral lines are marked by their symbols (see text).
The spectral line marked by * is

determined by phonon relaxation [41]) all scale with the square of the exciton wave function at $r=0$ [42], (see Eq. (1.1)), in a simple hydrogen model [25] the total intensity of an SHG process should scale with $S_0(n,P) \propto P^2/n^3$ where the proportionality constant is the same for all S states. However, as can be seen by comparing the initial values for n=5 and 6, this is not the case. They should differ at low powers by a factor of $(5/6)^3 \approx 0.58$, but are almost equal.

Besides the total intensity, spectral lines are characterized by their higher moments, the first giving the spectral position, a second the width of the line. Here we use a fitting of the lines by an asymmetric sech-function

$$D\left(E,E_0,S_0,s_1,s_2\right)=\frac{S_0}{s_1+s_2}\left\{\operatorname{sech}\left(E,E_0,s_1\right)\Theta\left(E_0-E\right)+\operatorname{sech}\left(E,E_0,s_2\right)\Theta\left(E-E_0\right)\right\}. \quad (1.4)$$

These parameters are plotted in Fig. 6 versus pump power for all S excitons with n=3 to 7. Here, the points represent the experimental results, while the full lines are a fit with relations derived from the simple absorption-followed-by-emission description.

The energy shifts of the lines $\Delta E$ are shown in panel a). If exciton-exciton interaction is the reason for the shift, it should scale as the density, i.e., the square of the pump power. However, as is exemplified by the full lines, it instead follows the square root of density, as represented by the relation $\Delta E(P) \propto P$. This shows clearly that a simple AfE description of resonant SHG is not possible. Also, these shifts saturate at higher pump powers. For the line broadening, we clearly observe an asymmetry, in that the low energy width $s_1$ increases much more than the high energy width (Fig. 6b). Note that the broadenings do not saturate, but seem to increase more at higher pump powers. This seems to be related to the existence of an additional relaxation channel at high powers, i.e., a shortening of the lifetime, as already guessed from in the peak area in Fig. 4.

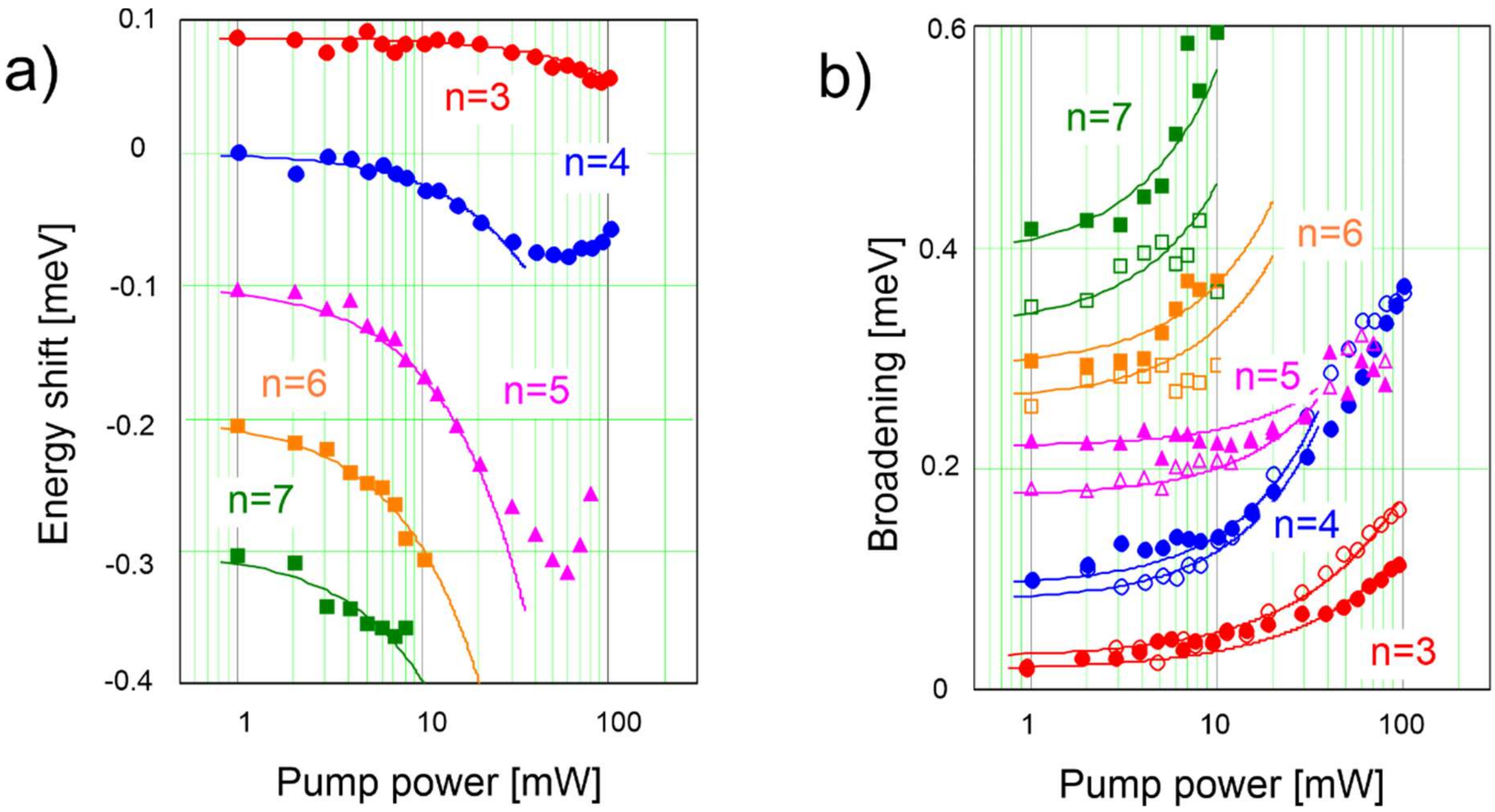

Fig. 6: Overview of the spectral shifts and widths obtained by fitting the "pure" exciton SHG spectra with a single double-sided SECH-line (Eq. (1.4)) representing the relevant transition to the nS state. Panel a) shows the relative energy position $\hbar\omega_0 - \hbar\omega_{nS}$, and panel b) the broadening parameters $s_1, s_2$. The data (symbols) are plotted versus pump power (logarithmic scale) for states from $n = 3$ to $n = 7$. Full symbols give the low energy, open symbols the high energy side of the spectral line. The full lines in panel a) show a fit assuming a linear dependence on pump

One should keep in mind that all these qualitative dependences do not allow to derive directly the interaction parameters, as they represent a complex average over the spatial and temporal dependent exciton distribution.

From Fig. 3 it becomes clear that, especially at higher powers and for large principal quantum numbers the spectra consist of many lines, albeit much weaker than the main resonance, due to energetically neighbouring exciton states (a complete list of the relevant exciton states is given in Table II). In order to obtain the "true", i.e., the undistorted line shape of the dominant line, one should subtract these lines from the spectrum. To include the possibility that the lines become asymmetric (see, e.g., the 6S line at 10 mW in Fig. 2d), we fitted the complete spectrum with a superposition of asymmetric sech lines

$$S(E) = \sum_k D\left(E, E_{0k}, S_{0k}, s_{1k}, s_{2k}\right) \quad (1.5)$$

and subtracted from the measured spectrum the residuum, i.e. the sum in Eq. (1.5) except the dominant term. Typical examples of this procedure are shown for the 5S measurement in Fig. 5. As additional lines we have to consider the $4D_2$, $5D_1$, $5D_2$, and the 6S lines. By this procedure we clearly obtained single resonance lines without much lesser interdependence with other exciton states. These spectra will then be compared in Section IV with the results of the theory of resonant SHG developed in section III. By this we will get a full understanding of the line shape, the line asymmetry at high powers, and the peculiar dependence of the peak area with quantum number and are able to derive quantitative values for the exciton interaction.

## III. Theory of resonant SHG from an interacting exciton system

### A) General considerations

The theoretical description of the SHG process is long established both in a classical and a quantum picture [29,31]. Here we use a semiclassical theory, whereby the electromagnetic fields are assumed to be classical, i.e., coherent, while the states of matter have to be described quantum mechanically.

While for the case of nonresonant SHG the electromagnetic fields obey the usual coupled Maxwell wave equations [31], those cannot be applied in the resonant case. To derive the coupled equations in this case, one has to start (see e.g., [29]) from a system Hamiltonian that considers besides the pump and SHG field two types of exciton states (i) the intermediate states $Y_\alpha$ with energies much larger than the pump photon energy $\hbar\nu$, which are dipole coupled to the pump field and (ii) the final target states $X_n$, which are resonant with two times pump photon energy and couple via a quadrupolar interaction to the SHG field. Further, we assume that the fields can be described classically, while the excitons are considered as bosons.

$$
\begin{aligned}
\hat{h}(\mathbf{r})/\hbar =& \sum_n \omega_n X_n^\dagger(\mathbf{r},t)X_n(\mathbf{r},t) + \sum_\alpha \nu_\alpha Y_\alpha^\dagger(\mathbf{r},t)Y_\alpha(\mathbf{r},t) \\
&+\sum_\alpha g^{0\to\alpha} Y_\alpha^\dagger(\mathbf{r},t)E_{IR}(\mathbf{r},t)e^{-i(\nu t-k(\nu)z)} + g^{\alpha\to 0}Y_\alpha(\mathbf{r},t)E^*_{IR}(\mathbf{r},t)e^{i(\nu t-k(\nu)z)} \\
&+\sum_{\alpha,n} g^{\alpha\to n} X_n^\dagger(\mathbf{r},t)Y_\alpha(\mathbf{r},t)E_{IR}(\mathbf{r},t)e^{-i(\nu t-k(\nu)z)} + g^{n\to\alpha}X_n(\mathbf{r},t)Y_\alpha^\dagger(\mathbf{r},t)E^*_{IR}(\mathbf{r},t)e^{i(\nu t-k(\nu)z)} \\
&+\sum_n q^{n\to 0} X_n(\mathbf{r},t)E^*_{\mathrm{SHG}}(\mathbf{r},t)e^{i(2\nu t-k(2\nu)z)} + q^{0\to n}X_n^\dagger(\mathbf{r},t)E_{\mathrm{SHG}}(\mathbf{r},t)e^{-i(2\nu t-k(2\nu)z)} \\
&+\sum_m \int d\mathbf{r}' V_{nm}(\mathbf{r}-\mathbf{r}')\hat{X}_m^\dagger(\mathbf{r}',t)\hat{X}_m(\mathbf{r}',t)X_n^\dagger(\mathbf{r},t)\hat{X}_n(\mathbf{r},t)
\end{aligned}
\tag{1.6}
$$

From the Hamiltonian we can derive the following equation of motion for the exciton amplitude

$$
\begin{aligned}
\frac{\partial}{\partial t}\hat{X}_n(\mathbf{r},t) =& -\left(i\delta_n + \frac{\Gamma_n}{2}\right)\hat{X}_n(\mathbf{r},t) + iG_{SHG,n}E_{IR}^2(t) - iq_n\hat{E}_{SHG}(t)\exp\left(-i\Delta k_n z\right) - \\
&\sum_m i\int d\mathbf{r}' V_{nm}(\mathbf{r}-\mathbf{r}')\hat{X}_m^\dagger(\mathbf{r}',t)\hat{X}_m(\mathbf{r}',t)\hat{X}_n(\mathbf{r},t)
\end{aligned}
, \tag{1.7}
$$

with $G_{SHG,n} = \sum g^{0\to\alpha}g^{\alpha\to n}/\Delta_\alpha$ and $2/\Gamma_n$ the decay time of the exciton amplitude and $\delta_n = E_{nX}/\hbar - 2\omega_{IR}$ the frequency detuning, where $\omega_{IR}$ is the frequency of the IR pump laser .

In the last term on the right-hand side of Eq. (1.7), $V$ denotes the interaction potential between two excitons (divided by $\hbar$), and as we neglect higher correlations, the model is restricted to the limit of low densities. Hereby, the real part of the integral acts like an additional frequency shift, while the imaginary part would lead to an additional decay of the exciton amplitude. We further assume that the three-point correlation function in the interaction term can be factorized, i.e., $\left\langle \hat{X}_m^\dagger(\mathbf{r}',t)\hat{X}_m(\mathbf{r}',t)\hat{X}_n(\mathbf{r},t)\right\rangle = \rho_X(\mathbf{r}',t)\left\langle X(\mathbf{r},t)\right\rangle$ giving for the interaction term

$$
\int d\mathbf{r}' V_{nm}(\mathbf{r}-\mathbf{r}')\left\langle \hat{X}_m^\dagger(\mathbf{r}',t)\hat{X}_m(\mathbf{r}',t)\hat{X}_n(\mathbf{r},t)\right\rangle = \int d\mathbf{r}' V_{nm}(\mathbf{r}-\mathbf{r}')\rho_{mX}(\mathbf{r}',t)\left\langle \hat{X}_n(\mathbf{r},t)\right\rangle \tag{1.8}
$$

and leads to a formal closure of the set of equations of motion, but still requires the solution of a set of integro-differential equations.

For the SHG field one gets

$$
\left(\frac{\partial}{\partial z} + \frac{n_b}{c_0}\frac{\partial}{\partial t}\right)E_{SHG}(\mathbf{r},t) = -\sum_n i\frac{n_b}{c_0}q_n^*\left\langle \hat{X}_n(\mathbf{r},t)\right\rangle \exp\left(i\Delta k_n z\right) \tag{1.9}
$$

with $\left\langle \hat{X}_n(\mathbf{r},t) \right\rangle$ the average of the exciton amplitude and $q_n$ the quadrupolar coupling strength. The refractive index at the exciton resonance is denoted by $n_b$. The IR field itself obeys the equation

$$\left( \frac{\partial}{\partial z} + \frac{n_{IR}}{c_0} \frac{\partial}{\partial t} \right) \hat{E}_{IR}(\mathbf{r},t) = i \frac{n_{IR}}{c_0} \sum_n \frac{2iG_{nSHG}^2}{\Gamma_n / 2 + i\tilde{\delta}_n} \left| E_{IR}(\mathbf{r},t) \right|^2 E_{IR}(\mathbf{r},t) \quad , \tag{1.10}$$

from which one can derive the corresponding equation for the intensity $I_{\mathrm{IR}} = \left| E_{IR} \right|^2$

$$\left( \frac{\partial}{\partial z} + \frac{n_{IR}}{c_0} \frac{\partial}{\partial t} \right) I_{IR}(\mathbf{r},t) = -\alpha_2 I_{IR}(\mathbf{r},t)^2 \ , \tag{1.11}$$

which is equivalent to the usual equation for two-photon absorption [31] with an absorption coefficient $\alpha_2 = \frac{n_{IR}}{c_0} \sum_n 2\Gamma_n G_{nSHG}^2 / \left( \Gamma_n^2 / 4 + \tilde{\delta}_n^2 \right)$. Here we included the interaction term in an effective detuning

$$\tilde{\delta}_n = \delta_n + \sum_m \int d\mathbf{r}' V_{nm}(\mathbf{r}-\mathbf{r}') \rho_{mX}(\mathbf{r}',t) \ . \tag{1.12}$$

Obviously, this leads to a reduction of the absorption coefficient with increasing exciton density, which is commonly known as the Rydberg blockade. This buildup of density correlations can be interpreted as an excluded volume that is no longer available for the absorption process.

In addition, we included the wave vector mismatch $\Delta k_n = k_{n,opt} - 2k_{IR}$. Here $k_{n,opt} = n_b E_{nX} / (\hbar c)$ is the wave vector of light at the exciton resonance and $k_{IR}$ the wave vector of the IR laser beam. Actually, the mismatch is quite large, as $2k_{IR} = 29.1\ \mu\mathrm{m}^{-1}$ and $k_{n,opt} = 33.6\ \mu\mathrm{m}^{-1}$ leading to a spatial oscillatory behavior of the SHG light with a period of the interaction length $L_c = 2\pi / \Delta k = 1.4\ \mu\mathrm{m}$. Note, that because of the phase mismatch, the outcoming field strength depends crucially on the thickness of the sample and due to the dispersion of the refractive index on the wavelength of the pump pulse.

We stress that we consider the coherent limit, i.e., neglect higher order correlations of the exciton amplitudes, so that the exciton density is given by $\rho_{nX}(\mathbf{r},t) = \left| \left\langle X_n(\mathbf{r},t) \right\rangle \right|^2$. In this model RSHG is a fully coherent process and $2/\Gamma_n$ corresponds to the phase decoherence time of a two-level system $T_2$ and the exciton lifetime $T_1$ is just $T_2/2$.

For thin samples with a thickness of a few interaction lengths, one can neglect any depletion of the pump laser, setting $E_{IR}(t) = \text{constant}$, while for thick crystals one has to solve the full set of equations. This is required, e.g., for the measurements of two-photon absorption cross sections.

To be able to straightforwardly compare the results of the calculations with experiments, we use the following units: (i) time in ps, (ii) exciton density in $1/\mu\text{m}^3$, which means that the exciton amplitude is given in $1/\mu\text{m}^{3/2}$. The electromagnetic fields are expressed in units of $E_0 = \sqrt{\hbar\omega/(2\varepsilon_0 n_b^2)}$, which gives them a dimension also of $1/\mu\text{m}^{3/2}$. The relation of light intensity, i.e. $I = |E|^2$ to the measurable physical intensity is then simply $I_{\text{phys}} = \frac{1}{4}\hbar\omega\frac{c_0}{n_{IR}}I$.

In the case of a coherent process the time-averaged spectrum of the emitted light is simple given by the absolute square of the Fourier transform of the emitted SHG field [35], which is convolved with a Gaussian (halfwidth $60\ \mu\text{eV}$) to take the finite spectral resolution of the spectrometer into account.

As the whole process depends nonlinearly on the IR power, the spatial profile of the laser pulse has to be considered. Assuming radial symmetry in the transversal plane, we integrated the solution over the gaussian mode profile with a 16-point Simpson routine, which was sufficiently accurate.

B) Exciton-Exciton interactions

To further advance in the considerations, we have to specify details of the interaction potential. As is well-known from atomic physics, there are two regions for the description of interatomic interactions, which are separated by the so-called Le Roy radius $R_{LR}(n,l) = 4(<r^2>)^{1/2}$ [43], where $<r^2>$ is the mean square of the electron-hole distance. Using scaled hydrogen wave functions for the exciton states, we obtain $\sqrt{<r^2>} = 2n\sqrt{5n^2-1-3(L+1)}a_B$ with the exciton Bohr radius $a_B = 1.10$ nm [5]. From this we get for the Le Roy radius of the 3S state 42.3 nm, while that for the 7S state it would be 239 nm. For average distances larger than this radius, or equivalently densities much smaller than a critical density $\rho_{LR} = 3/(4\pi R_{LR}^3)$, which would be $1053\ \mu\text{m}^{-3}$ for the 3S and $17.1\ \mu\text{m}^{-3}$ for the 7S state (see Table I), one can neglect higher order effects like higher multipole orders or exchange interactions. If only a single exciton state is excited in the crystal one can use a simple van-der Waals-like description with the usual inverse power law for $V(\mathbf{r})$

$$V(\mathbf{r}) = \frac{C_p/\hbar}{r^p} \tag{1.13}$$

with $p = 6$ for an exact vdW interaction. This was recently confirmed also for excitons [6]. In this work the interaction laws for excitons by the induced dipole-dipole interaction have also been calculated in the heavy hole limit for states $n \geq 5$. The results for the S-S interactions can be summarized in the following expression

$$\begin{gathered} C_{6S}(n) = F_6(n) n^{11} \quad \text{with} \quad F_6(n) = f_0 + f_1 n^{f_2} \\ f_0 = -1.376 \cdot 10^6 \text{ nm}^6 \mu\text{eV} \, ; f_1 = 2.2057 \cdot 10^6 \text{ nm}^6 \mu\text{eV} \, ; f_2 = \text{-}0.8313 \end{gathered} \quad . \tag{1.14}$$

However, we face the problem that the integral over the interaction potential diverges for $r \to 0$. This means that the system builds up correlations such that around an exciton there is an exciton-free volume, see e.g., Ref. 7. We can take care of this "blockade effect", which commonly is called "Rydberg blockade" by excluding in the integration this volume, given by a critical radius $R_c$. To obtain a criterion for it, we have a closer look at the equation of motion (1.7). Here we see that the interaction term gives rise to an additional detuning, i.e., to oscillations of the exciton amplitude. If the period of this oscillation is shorter than the pulse width, then it will average out the effect of the interaction term and we can neglect it. Obviously, this takes place if $\Delta t_P V(R_c) \geq 1$. This gives the following condition for $R_c$

$$R_c(n) = \left( \frac{|C_p(n)|}{\hbar(1/\Delta t_P + \Gamma_X/2)} \right)^{1/p} , \tag{1.15}$$

where we have taken the finite decay time into account (see Appendix B).

Furthermore, due to the $r^{-p}$ dependence of the potential, the overwhelming contribution to the integral comes from the nearest surrounding of the exciton. If the excitation spot is much larger than $R_c$, one can neglect the spatial variation of the density and perform the integration giving (for $p > 3$)

$$\int d\mathbf{r}' V(\mathbf{r} - \mathbf{r}') \rho_X(\mathbf{r}', t) = a(n) \rho_X(t) \tag{1.16}$$

with

$$a(n) = 4\pi\, sign(C_p)\frac{1}{(p-3)}\left(\hbar / \Delta t_P + \frac{1}{2}\hbar\Gamma(n)\right)^{(p-3)/p} |C_p(n)|^{3/p} \; . \tag{1.17}$$

From this follows a simple relation between interaction strength and blockade volume

$$a(n) = \frac{3}{p-3}\mathrm{sign}(C(n))\left(\hbar / \Delta t_P + \frac{1}{2}\hbar\Gamma(n)\right)V_{BL}(n) \; . \tag{1.18}$$

The simple van der Waals law then predicts a dependence of the interaction constant on principal quantum number as

$$a(n) \propto \sqrt{F_6(n)}n^{5.5} \quad . \tag{1.19}$$

## IV. Comparison of experimental results with theoretical calculations

For the analysis of the experimental spectra we solved the coupled system of equations derived in section III. For the numerical calculations we used the program routines provided by the MATHCAD package. To give quantitative results, we use the following parameters:

a) Damping
   Here we use the results of one-photon absorption experiments, which gave for the linewidth of nS excitons $1/\Gamma = 0.094\ \mathrm{ps}\cdot n^3$ [5]. For *n=3* to *5* this agrees well with the results from direct lifetime measurements [33].

b) SHG coupling to the fundamental IR light. This can be determined from the 2PA experiments (see Appendix B) giving a coupling constant for the S states $G_{SHG}(n) = (4.604 \pm 0.6)10^{-5} / n^{3/2}\ \mu\mathrm{m}^{-3/2}\mathrm{ps}^{-1}$. For states with higher principal quantum number we assumed a dependence of the SHG absorption constant $\propto 1/n^3$ as found in [25].

c) Quadrupole coupling. This can be obtained through measurements of the one-photon absorption in electric fields [5,44]. The analysis gives a law $q_{nS}(n) = (4.94 \pm 0.8)n^{-3/2}\mathrm{ps}^{-1}$, while the coupling constant for the $D_2$ states is $q_{nD2}(n) = (2.51 \pm 0.5)n^{3/2}\mathrm{ps}^{-1}$.

d) To calculate the wave vector mismatch, we used the following expression for the refractive index $n(E_{phot})^2 = 4.7317 + 2.00987/(1 + 0.12282(E_{phot}/eV)^2)$ [45], where $E_{phot}$ denotes the photon energy.

The remaining free parameters are the exciton-exciton interaction constants $a_{Sn}$, which were adjusted to obtain the best agreement of the calculated with the experimental spectra. The blockade volumes $V_{BLn}$ are directly related to the interaction constants by Eq. (1.32). As the latter would influence the two-photon absorption (2PA), one can in principle use these results for a check of consistency. Unfortunately, we were able to measure 2PA only for *nS=3* (see Appendix B), so we can only use these values.

Before we discuss these results, we give an overview of the general results of the model, especially with respect to the pump power range, where we expect the solution to give meaningful results. For this we assumed a homogeneous excitation, resulting in a spatially constant exciton densities, which are shown in Fig. 7. As exciton-exciton interaction constants we assumed $a_{Sn}(n) = 2.0 \cdot 10^{-8} \cdot n^{7.5}$, while the blockade volume is given by Eq. (1.18), which comes close to the fit values.

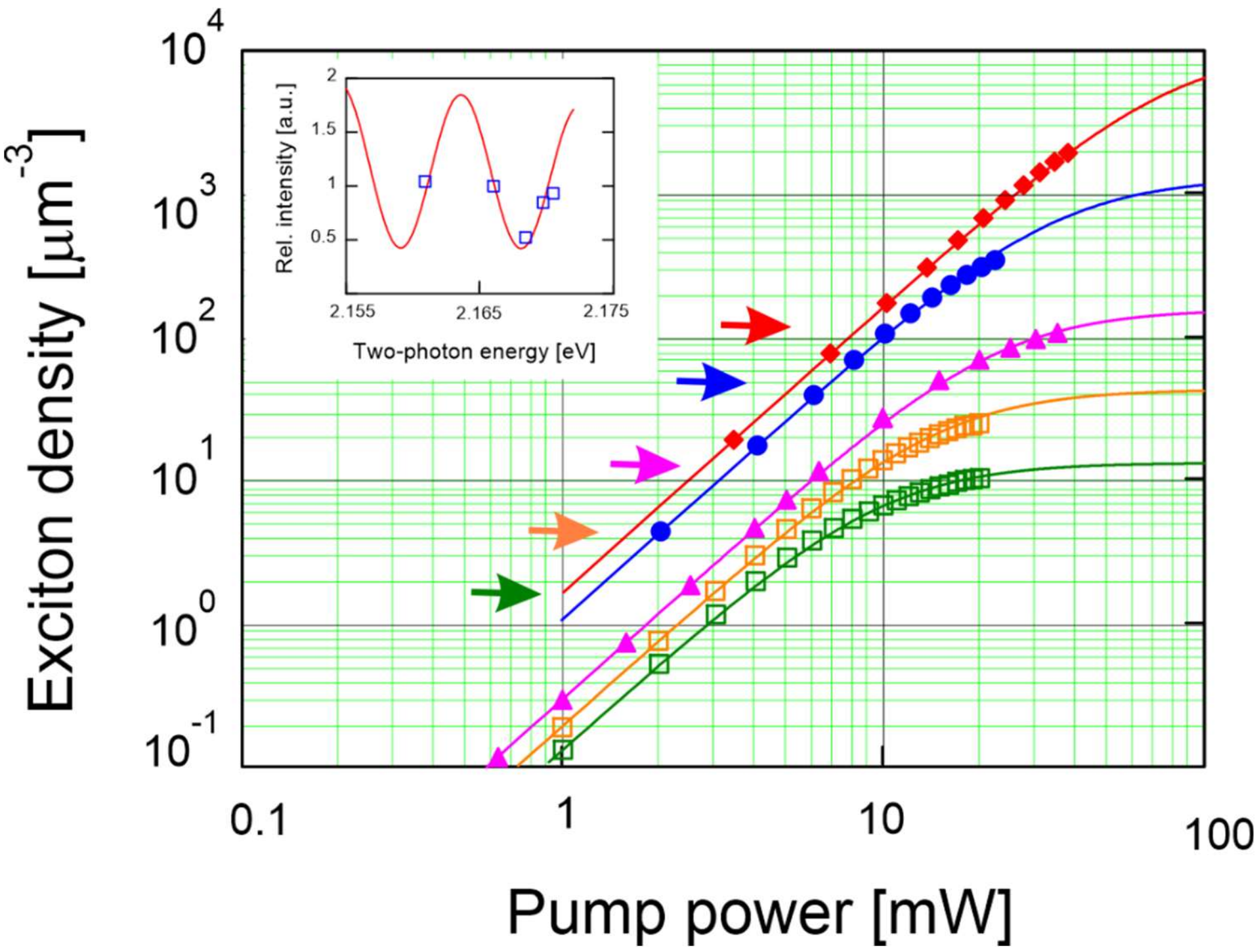

Fig. 7: Calculated densities of excitons with principal quantum numbers from n=3 to 7 as function of IR laser power. Points are results of the calculation, the full lines give an approximation by the simple saturation law (Eq. 1.2). The arrows mark one tenth of the critical Le Roy densities given in Table I, which we take as an upper bound for the validity of our model. This results in a power limit of about 10 mW, almost independent of quantum number n. The inset shows the relative intensities of RSHG as function of two-photon energy for a crystal of thickness of 29 µm . The full line gives the results of the calculation showing the typical oscillatory behaviour of a coherent SHG, the blue square points are the ratios of measured RSHG signals to those expected from Eq. (1.1).

We clearly see, that the power range where we expect the theoretical model to be valid is restricted to powers below 10 mW independent of the quantum number. Indeed, this agrees well with the results of the fitting procedure as shown in Fig. 8, where deviations occur above this power range (compare the data for *n=4* and 20 mW (magenta points in panel a) of Fig. 8).

In the valid power range, the experimentally obtained “pure” resonance spectra can be fitted extremely well by our model reproducing especially the asymmetric sech-like line shape and the anomalous behavior of the intensity on principal quantum number of the exciton states as due to the wavelength dependence of the wavevector mismatch (see inset in Fig. 7). This agreement substantiates the coherent character of the RSHG process, as a simple AfE mechanism would not show such a behavior.

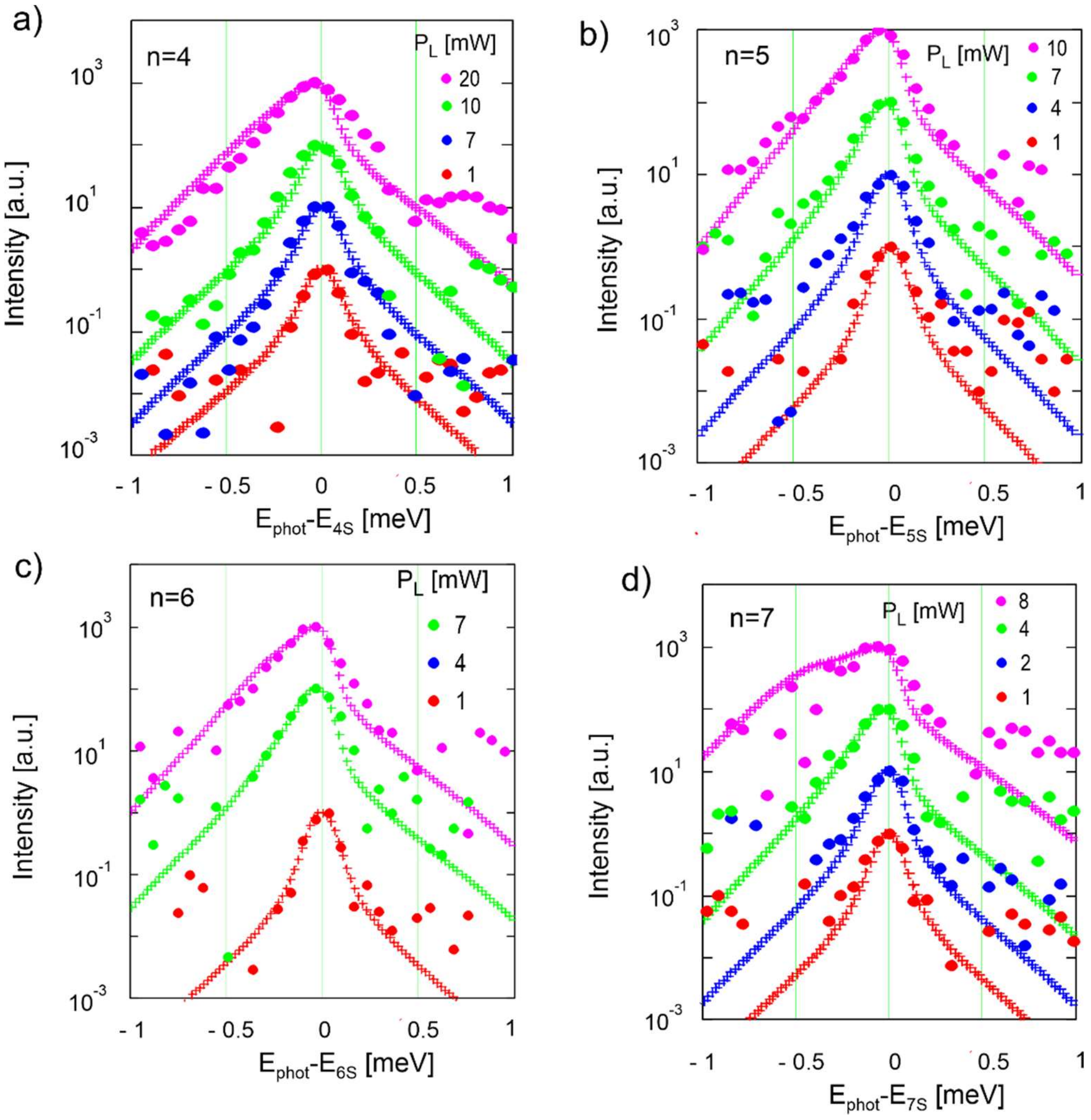

Fig. 8: Comparison of the corrected “pure” spectra (points) with the results of the theoretical model (full lines) for principal quantum numbers from n=4 to n=7 in panels a) to d) at the pump powers given in the panels.

The overall excellent agreement between theory and experiment allows to determine the interaction constants quite accurately with an error of less than 30%, which is mainly given by the inaccuracy of the two-photon interaction constant. The values are given in Table I and also shown in Fig. 9 together with those given by the standard atomic-like van der Waals model.

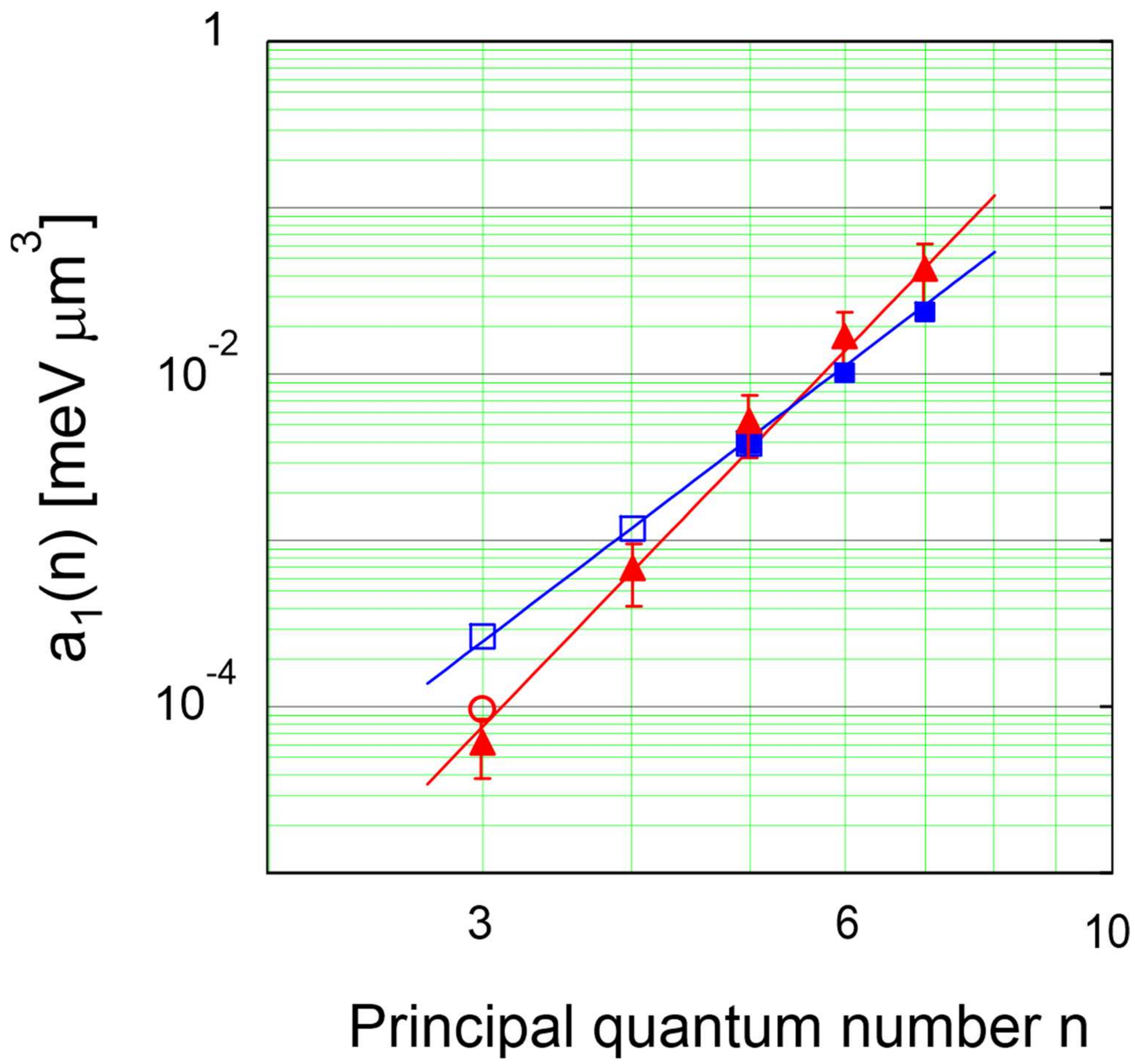

Fig. 9: Comparison of exciton interaction strengths $a_1(n)$ obtained from the experiments with results of an atomic-like van der Waals theory. The filled triangles show the results obtained by the analysis described in Section IV, the open circle for n=3 is calculated from the blockade volume by Eq. (1.32), the full line is a power law $c_a \cdot n^{7.5}$ with $c_a = 2.0 \cdot 10^{-8}\,\mathrm{meV\mu m^3}$ . The blue full squares are calculated using the $C_6$ interaction constants (for $n \geq 5$) taken from Ref. 5 (see Eq. 1.14), the open square is an extrapolation of this equation, the full line is the power law $c_{\mathrm{vdW}} \cdot n^{5.5}$ with $c_{\mathrm{vdW}} = 5.89 \cdot 10^{-7}\,\mathrm{meV\mu m}$ expected for the atomic-like van der Waals interaction (see Eq. 1.16).

## V. Discussion

The main result of this work is the absolute determination of the exciton-exciton interaction for excitons in high principal quantum number states up to *n=7*, which extents in the range of Rydberg excitons. While the values agree in their magnitude within a factor of two with the results of an atomic-like van der Waals theory [6], they follow a somewhat different scaling law with the principal quantum number, instead of the expected proportionality $\propto n^{5.5}$ we find a scaling with $\propto n^{7.5}$. This difference can be due to a number of reasons. The first is that the difference might be related to the atomic-like van der Waals theory itself, as the calculations were performed in the limit of infinite center-of-gravity mass for the excitons. For the yellow states in $Cu_2O$ this approximation is clearly inappropriate, as the masses of both electron and hole are almost equal to the free electron mass [5]. Other influences might come from the fact that excitons are existing in a solid-state environment (see Fig. 1), with strong influences especially from an electron-hole plasma [18,19,21] or charged impurities [46]. The concentration of the latter can be estimated from the analysis of high resolution transmission experiments of the yellow Rydberg states [46]. Here, the samples used in the experiments are of high quality with maximal principal quantum numbers up to *n=19* to *20* [47], which proves a charged impurity concentration below $10^9\,\mathrm{cm}^{-3}$ with negligible influence on Rydberg states with $n<10$ [46]. On the other hand, we have seen from the analysis of the nonlinear absorption (see Appendix B) that at high pump powers we excite a significant concentration of an electron-hole plasma due to three-photon absorption in the $\Gamma_8^-$ conduction band and holes in the $\Gamma_{7,8}^+$ valence bands (blue-violet plasma), which interacts with the Rydberg states. At a power of 3 mW we estimate a concentration of $\rho_{avgEHP} = 0.1/\mu\mathrm{m}^3$ (see Appendix B). As explained in detail in Ref. 19, there are two main effects of an EHP, (i) additional broadening of the exciton lines by plasmon scattering, and (ii) vanishing of exciton states due to the Mott effect. Obviously, we do not observe any Mott effect, as the exciton lines are still present up to the highest powers, so we need to consider only the plasma broadening. According to [40] for excitons in $Cu_2O$ it depends on the density $\rho_{\mathrm{pl}}$ and temperature $T_{\mathrm{pl}}$ and the principal quantum number as

$$\Gamma_{\mathrm{pl}}(n) = C_{\mathrm{pl}}(n-1)^{3.3}\frac{\rho_{\mathrm{pl}}}{\sqrt{T_{pl}}} \tag{1.20}$$

with a constant $C_{\mathrm{pl}}$, that for P excitons is given by $C_{pl} = 3.2\,\mu\mathrm{eV}\mu\mathrm{m}^3\mathrm{K}^{-1/2}$. To apply this for S excitons we have to consider that here only anti-Stokes scattering from S to P states is possible reducing $C_{\mathrm{pl}}$ by a factor of five and giving an extra factor $exp(-\Delta E_{\mathrm{SP}}/k_{\mathrm{B}}T_{\mathrm{pl}})$ where $\Delta E_{\mathrm{SP}} = 29/n^3\,\mathrm{meV}$ is the energy difference of the *nS* and *nP* states [5]. While the plasma temperature will be around 10 K for sample temperatures below 2 K (see the discussion in [19]), the plasma density will scale with the third power of the pump intensity. Plasma effects will become important if the plasma scattering rate is larger than the exciton decay rate. For the 5S state, e.g., this requires a density of the order of $5\,\mu\mathrm{m}^{-3}$ corresponding to a pump power of 10 mW, which is just the power where the effects of the additional scattering mechanism postulated above sets in for the 5S state (see arrows for the other states in Fig. 4).

To include the effect of a plasma in the theory we have to augment the exciton equation of motion (Eq. (1.7)) by a term

$$-g_{pl} \cdot (n-1)^{3.3} \left|X_n(\mathbf{r},t)\right|^6 \hat{X}_n(\mathbf{r},t) \;, \tag{1.21}$$

where $g_{pl}$ comprises all proportionality constants. It is easy to see that this additional broadening not only leads to a reduction of SHG efficiency equivalent to that of Eq. (1.3) but allows to predict the dependence of the critical power $P_{sc}$ on quantum number to be given by $P_{sc}(n) \propto n^{1.1}$. For the states $nS = 4$ and $nS = 5$ this ratio is predicted as 0.78, while it experimentally was obtained to be 0.75, in excellent agreement.

From this we can conclude that in the power range where we applied our theoretical model the b/v plasma plays no role for the excitons so it does not influence our results.

If we look at the spectra of the $nS = 3$ exciton line, we see that it behaves differently, as at high powers the line stays almost symmetric, while it should show some asymmetry. In the same vein we stress that the exciton densities achieved in the experiments at high pump powers imply for the 3S state more than one exciton within a volume of the size of an optical wavelength (see Fig. 7). Thus, we highly suspect cooperative radiative processes to occur in these power regions, which are not included in our theoretical description.

## V. Conclusions

In this paper we have shown that by using SHG with ps laser pulses it is possible to study yellow excitons in $Cu_2O$ with principal quantum numbers up to $n=7$ at such densities that clear non-linear effects become observable. These consist primarily in (i) an energy shift of the exciton resonance lines, (ii) a broadening of the lines with a strong asymmetry and (iii) a saturation of the intensity, but the experimental results reveal already a very rich scenario of non-linear effects like, e.g., multi-exciton interactions. The detailed analysis of these effects, however, would go beyond the aims of this paper. In the same vein we stress that the exciton densities achieved in the experiments at high pump powers imply more than one exciton within a volume of the size of an optical wavelength. Thus, we highly suspect cooperative radiative processes to occur in these power regions.

In the experimental scenario, the presence of a yellow EHP can be excluded, but the occurrence of a blue-violet plasma excited by three-photon absorption at high pump power has to be considered, but was shown to be ineffective in the power range of the experiments and therefore could not explain the non-linear effects. Hence, these can only stem from exciton-exciton interactions. The observed energy shift of the lines scales at low pump powers linearly with the exciton density, which could be determined from measuring the two-photon absorption directly. The analysis allows us to deduce the strength of the exciton-exciton interaction in the S-S scattering channel for principal quantum numbers from 3 to 7 quantitatively. As in this power range the mean exciton distance is larger than the Le Roy radius, the exciton-exciton interaction should be determined simply by an interaction law. This allows to compare our results with the standard van der Waals theory [6]. To this end, we put forward a Rydberg blockade model, from which we can deduce the line shifts and the saturation behavior given the interaction law of the excitons. While we find the interaction strength to be of the same order as predicted by theory the scaling with the principal quantum number is different from atomic-like van der Waals calculations [6]. We can rule out that this is due to too high exciton densities, as we restricted the analysis to low pump powers, where the van der Waals approximation should be valid. One possible reason may lie in deficiencies of the theory itself, as the hitherto used energy levels and transition dipole moments were derived from theoretical exciton models, which assume an infinite translational exciton mass and not from experiment, i.e. by Terahertz-spectroscopy [48]. At higher exciton densities the inadequacy of a simple van der Waals description is already visible in our experiments in form of the observed nonlinearity of the energy shifts with power. This may point towards the claims of theoretical papers (see e.g. Refs. 49 and 50) that due to the composite fermion nature the interaction between excitons cannot be described by a potential function depending only on the distance of the centers of masses of the excitons, but must be non-local. In either case, our results will provide a solid test for any theory of interacting excitons to be developed, e.g., along the lines of Ref. 51. On the experimental side in the future, it is urgently needed to extend the investigated exciton states to higher principal quantum numbers to overlap with other types of measurements. Progress here will require the use of laser pulses with much smaller line width, preferentially optimized to the linewidth of the Rydberg states to allow selective excitation of single states. And last but not least, the quantitative determination of exciton densities from two-photon absorption must be improved also in view of higher states, which will be extremely challenging as very small effects of less than 0.1% absorbance are expected.

**Acknowledgements**

We thank Wolf-Dietrich Kraeft, University of Rostock, Germany, for helpful discussions. D.S. thanks the Deutsche Forschungsgemeinschaft for financial support (project number SE 2885/1-1), the Dortmund side acknowledges the support by the Deutsche Forschungsgemeinschaft through the International Collaborative Research Centre TRR160 (Project A8) and TRR 142 (Project A11).

## Appendix A: Nonresonant SHG and reconstruction of the IR laser pulse

Besides the resonant SHG, i.e., SHG which involves resonantly excited intermediate states, we also excite SHG which can be attributed to nonresonant processes. As usual, this can be described by an almost wavelength-independent second order susceptibility $\chi^{(2)}$, which is non-zero also in centrosymmetric crystals if one includes higher order multipoles in the theoretical model. As in standard SHG, the spectrum of this process is given by the absolute square of the autoconvolution of the electric field of the pump pulse $E_{IR}(\omega) = E_0(\omega)e^{i\varphi(\omega)}$ as $I_{SHG}(\omega) \propto |E_{SHG}(\omega)|^2 \propto \left|\int E_{IR}(\omega')E_{IR}(\omega-\omega')d\omega'\right|^2$. Therefore, the spectrum of the nonresonant SHG can easily measured by a standard setup using an appropriate non-linear crystal like BBO. A typical measurement, as was done routinely to ensure the quality of the laser pulse, is shown in Fig. A1. Obviously, the spectrum shows besides the central dominant peak strong sidebands and can be approximated by a rectangular pulse of width $T_P = 3.56\,\text{ps}$ as shown by the blue crosses. A better approximation is given by the green line in Fig. A1a, which assumes a phase jump of $\pi$ at each minimum of the field amplitude. The Fourier-transform of this function gives the square of the amplitude (red dots) and phase (blue triangles) of the time dependent electric field of the pulse as shown in Fig. A1b. Here the full line is a fit with a supergaussian $\propto exp(-(t/\sigma)^n)$ with n=4 and $\sigma = 2.5\,\mathrm{ps}$. As long as we are interested only in the intensity of the SHG, one can simulate the complete spectrum by a superposition of 10 Gaussians (black line in Fig. 2a).

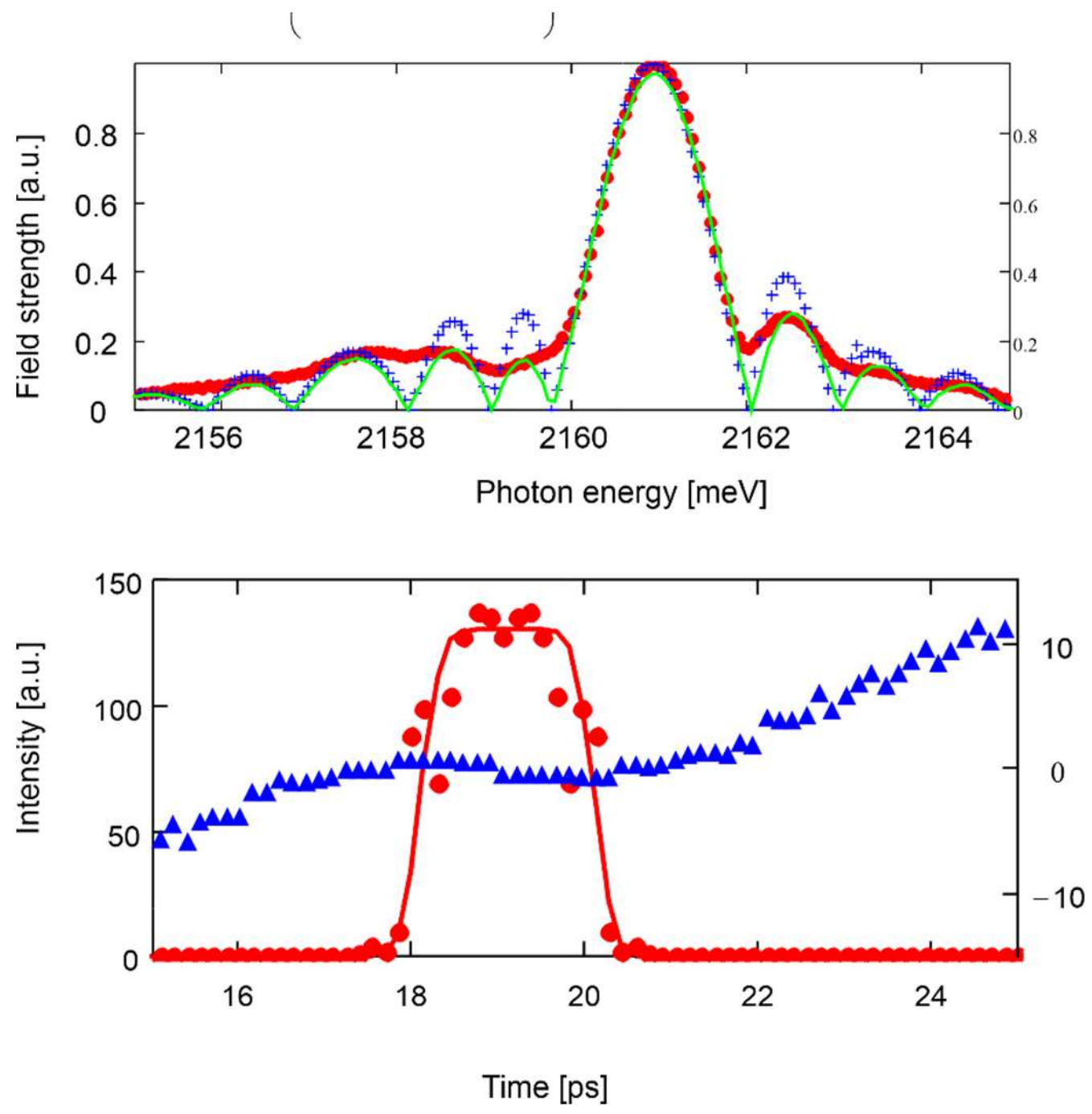

Fig.A1. The upper part shows the measured electric field spectrum of the second harmonic of the IR pump pulse (red dots, obtained using a BBO crystal) together with two model pulses. The FWHM of the central peak is 0.92 meV. The blue crosses show the spectrum of a simple rectangular pulse of width $t_P = 3.56$ ps given by a sinc(x)-function, while the full green line shows the actually used pulse spectrum, whereby the phase is assumed to jump by $\pi$ at each minimum of the field. The lower part shows intensity (red dots) and phase (blue triangles) as function of time of the field obtained by Fourier transforming the spectral field given by the green line in the upper part. The full line shows the pulse shape used in the theoretical calculations, a supergaussian $\propto exp(-(t/\sigma)^n)$ with n=4 and $\sigma = 2.5\,\text{ps}$.

**Appendix B: Determination of the two-photon coupling constant**

To obtain quantitative data for the exciton interactions, one has to know the density of the excitons. This can be determined by measuring the transmission loss of the IR pumping laser when its wavelength is in resonance with an exciton state. In contrast to nonresonant SHG, which as a $\chi^{(2)}$ process should not show 2PA [31], the excitation of a real exciton density is connected with an absorption process, as the excitons decay mostly nonradiative (see Eq. (1.10)). The setup is straightforward and uses a calibrated power meter to measure the power of the pump beam before $(P_{\mathrm{in}})$ and after the sample ( $P_{\mathrm{out}}$ ) outside the cryostat. To enlarge the transmission loss, we used for these measurements another crystal with a length of $L = 3.91\,\mathrm{mm}$. If $F_{\mathrm{rep}}$ denotes the repetition rate and $T_p$ the pulse duration of the laser pulse (we assume a rectangular shape with $T_p = 3.56\,\mathrm{ps}$ as a good approximation to the actual pulse) and a Gaussian spatial profile with beam waist $w_0$ the intensity at the entrance surface inside the sample is given by

$$I(r,z) = I_0 \left( \frac{w_0}{w(z)} \right)^2 \exp\left( \frac{-2r^2}{w(z)^2} \right). \tag{1.22}$$

The relation between the average laser power and the peak intensity is (note that we defined intensity as the photon fluence)

$$I_0 = \frac{2P_0}{\pi w_0^2} \frac{1}{T_P F_{\mathrm{rep}}} \frac{1}{E_{IR}} \text{ and } w(z) = w_0 \sqrt{1 + \left( \frac{z}{z_{\mathrm{R}}} \right)^2}. \tag{1.23}$$

Here we have $P_0 = T_W \left(1 - R_{IR}\right) P_{\mathrm{in}}$ ,whereby $T_W$ is the total transmission of the cryostat windows on one side and $R_{IR}$ denotes the reflectivity of the sample at the pump wavelength. The Rayleigh length is given by

$$z_{\mathrm{R}} = \frac{\pi w_0^2 n}{\lambda}. \tag{1.24}$$

The beam waist $w_0$ is related to the beam diameter $D_{\mathrm{FWHM}}$ by

$$w_0 = \frac{D_{\mathrm{FWHM}}}{\sqrt{2\ln 2}}. \tag{1.25}$$

The beam diameter has been measured as $D = (125 \pm 5)\ \mu\text{m}$ giving a $w_0 = 106\ \mu\text{m}$ and $z_R \approx 82\ \text{mm}$, so that variations of the beam diameter in the sample can be neglected.

If $T(I)$ denotes the internal transmission of the sample, then the pump power at the exit of the sample is

$$P(L) = E_{IR}\int_r I(r,L)rdr = \int_r T(I(r,0)I(r,0))rdr \tag{1.26}$$

so that the power outside the cryostat is $P_{\mathrm{out}} = T_W\left(1 - R_{IR}\right)P(L)$.

As example, we show the detailed evaluation in case of the 3S resonance. We get three sets of data by measuring the transmission, i.e., the ratio of $T = \frac{P_{\mathrm{out}}}{P_{\mathrm{in}}}$ for (i) without the crystal (calibration) at 1150.4 nm, (ii) with the crystal but at a wavelength well outside the 3S

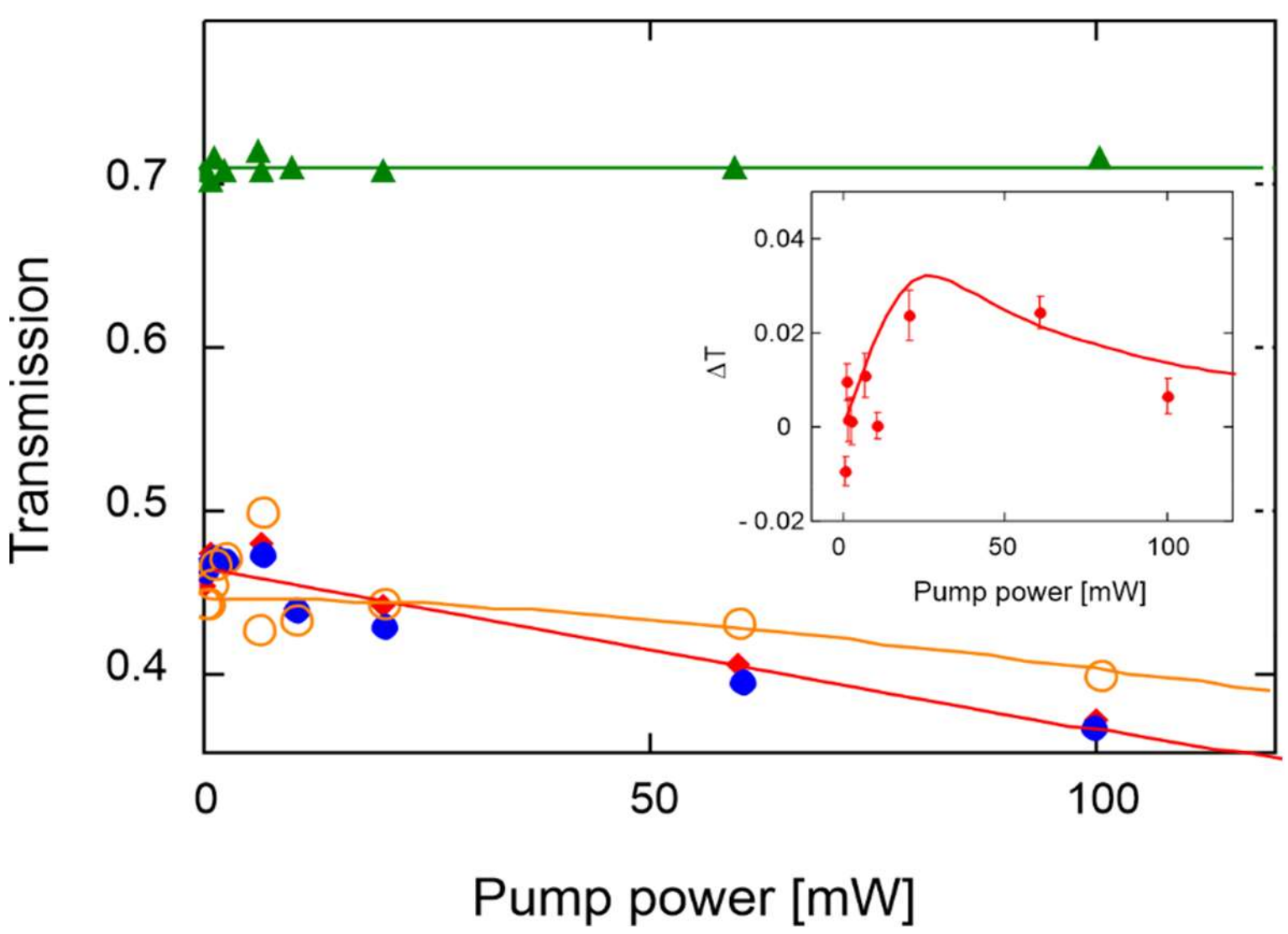

Fig. B1: Results of the transmission measurements of the 3S exciton. Green triangles: case (i), red full diamonds: case (ii), blue dots: case (iii), and ocher circles: case (iv). The errors are about 0.4%. The full green line gives the window transmission $T_w^2$, while the other full lines show the results of the modell for cases (ii) and (iv), see text for details. The inset shows the difference in internal transmission $\Delta T = T_{(ii)} - T_{(iii)}$. The red dots show the experimental results, the full line gives the fit with the model discussed in the text

resonance (at 1151.2 nm) and (iii) exactly at the 3S resonance at 1149.2 nm. The results are shown in Fig. B1 for laser powers from 0.1 mW to 100 mW.

While the transmission in case (i) is power independent as expected for quartz windows, they show a pronounced non-linear behavior with the $Cu_2O$ crystal, both in and out of the resonance, whereby the intensity loss is slightly larger in resonance. However, the loss outside the resonance needs as origin a further relaxation channel. This might be the background absorption due to the phonon-assisted absorption of the yellow 1S exciton in this region, which is quite strong (absorption coefficient $\alpha_1 = 18\ \text{mm}^{-1}$) or a three-photon absorption process into the blue and violet bands. To check this, we tuned the laser well outside the yellow absorption region ($E_{IR}$ = 1.012 eV, case (iv)) and measured also the transmission (ocher circles in Fig. B1). Indeed, we observed a measurable absorption, so that we have to include in all cases a three-photon process.

In a first step, one can estimate the exciton density already from the raw data using energy conservation, as each exciton is generated from two photons.

We have due to RSHG an additional loss of transmission of $\Delta T$ at an input power of $P_L$. This means that the energy loss per pulse is given by $\Delta E = \Delta T \cdot T_W \cdot (1 - R_{\text{IR}}) P_L / F_{rep}$ with $T_W \cdot (1 - R_{\text{IR}}) = 0.68$. This corresponds to a number of absorbed photons $\Delta N = \Delta E / E_{IR}$ and gives for the total number of excitons in the sample $\Delta N / 2$. Then the density of excitons is

$$\rho_{avg} = \frac{\Delta N}{\pi w_0^2 L} = 0.68 \frac{P_L \Delta T}{\pi w_0^2 F_{rep} E_{IR}} \quad . \tag{1.27}$$

The exciton densities calculated from the data of Fig. B1 are shown in Fig. B2 (squares). For 60 mW, e.g., this gives an average density of $\rho_{avg} = 1.4 \cdot 10^3\ \mu\text{m}^{-3}$.

In the same way, one can estimate the concentration of a blue/violet EHP by analysing case (iv). Here we have a transmission loss at 60 mW of $\Delta T = 0.031$ corresponding to an average density of the b/v-EHP of $\rho_{avgEHP} = 1.1 \cdot 10^3 / \mu\text{m}^3$. At a pump power of 3mW this would result in a concentration of $\rho_{avgEHP} = 0.1\,\mu\text{m}^{-3}$.

The equations (1.7) to (1.10) allow for a consistent quantitative modelling of the two-photon absorption. To include a possible three-photon absorption (3PA) it is useful to switch over to intensities, i.e., $I(\mathbf{r},t) = |E(\mathbf{r},t)|^2$. Then Eq. (1.10) gives the usual equation for 2PA [31]

$$\left(\frac{\partial}{\partial z}+\frac{n_{IR}}{c_0}\frac{\partial}{\partial t}\right)I_{IR}(\mathbf{r},t)=-\alpha_2 I_{IR}(\mathbf{r},t)^2-\alpha_3 I_{IR}(\mathbf{r},t)^3 \quad (1.28)$$

with $\alpha_2$ including both nonresonant and resonant 2PA and a 3PA term with $\alpha_3$ being the absorption coefficient. $\alpha_2$ is given by

$$\alpha_2=\alpha_{2\mathrm{nr}}+\alpha_{2\mathrm{res}}=\alpha_{2\mathrm{nr}}+\frac{2\Gamma_n G_{nSHG}^2}{\Gamma_n^2/4+\tilde{\delta}_n^2} \quad . \quad (1.29)$$

To account for the Rydberg blockade we approximate the effective interaction integral by Eq. (1.16) and obtain by assuming an adiabatic following of the exciton amplitude to the pump pulse

$$\alpha_{2\mathrm{res}}=\frac{n_{IR}}{c_0}\frac{2\Gamma_n G_{nSHG}^2}{\Gamma_n^2/4+\tilde{\delta}_n^2}=\frac{n_{IR}}{c_0}\frac{8G_{nSHG}^2}{\Gamma_n}\frac{1}{1+\left(V_{BL,n}\rho_{X,n}\right)^2} \quad , \quad (1.30)$$

whereby the blockade volume can be expressed as

$$V_{BL,n}=2a(n)/\Gamma_n \, . \quad (1.31)$$

Comparing with the expression derived in Section IV, where we did explicitly take the finite pulse width into account, one should interpolate both expressions by

$$V_{BL}(n)\approx a(n)\frac{1}{\Gamma_X/2+1/\Delta t_P} \quad . \quad (1.32)$$

As the generation of excitons is a local process we can describe it by the simple equation $\partial\rho/\partial t=\Gamma_{loss}I_{IR}(t)/2-\Gamma_X\rho$. Comparing with Eq. (1.28), we see that the generation rate is given by $g_X=\frac{c_0}{2n_{\mathrm{IR}}}\alpha_{2\mathrm{res}}$. The exciton density is then

$$\rho_X(I_{IR})=\frac{g_X}{2\Gamma_X}I_{IR}^2 \quad . \quad (1.33)$$

For the modelling, we start with case (iv), which turns out to be solely due to 3PA as can be seen by the ocher line in Fig. B1, which gives a fit with $\alpha_3=1.784\cdot10^{-18}$ $\mu\mathrm{m}^5$ and shows the characteristic negative curvature of the power dependence. Obviously, the 3PA is not negligible and we have to include it in the analysis of the 3S absorption. Thereby we can assume that the 3PA coefficient does only slightly change going from 1.012 eV to 1.08 eV as the three-photon energy is about 0.5 eV above the blue/violet bands. Then the analysis of case

(ii) requires additionally an effective 2PA coefficient of $\alpha_{nr} = 1.38 \cdot 10^{-11}\mu\text{m}^2$ (see full red line in Fig. B1). To model the resonant case (iii), we need a resonance absorption coefficient $\alpha_{2,3S} = 1.43 \cdot 10^{-11}\mu\text{m}^2$ and a blockade volume of $V_{\text{BL}} = 4.5 \cdot 10^{-4}\mu\text{m}^3$, which gives the full red line in the inset of Fig. B1. Using for $\Gamma_X = 0.384/\text{ps}$, we obtain for the 2PA coupling constant $G_{3\text{SHG}} = 8.86 \cdot 10^{-6}\mu\text{m}^{3/2}\text{ps}^{-1}$ and for the exciton interaction constant $a_S(3) = 8.86 \cdot 10^{-5}\mu\text{m/ps}$. The estimated error of these values is 30%.

A comparison of the densities obtained from the simple transmission analysis (see Eq. (1.27), open square symbols) and the results using Eq. (1.33) (full line) is shown in Fig. B2. While the overall dependence on pump power is the same, they differ by about 30%, which seems reasonable because of the rather crude estimate of the first method.

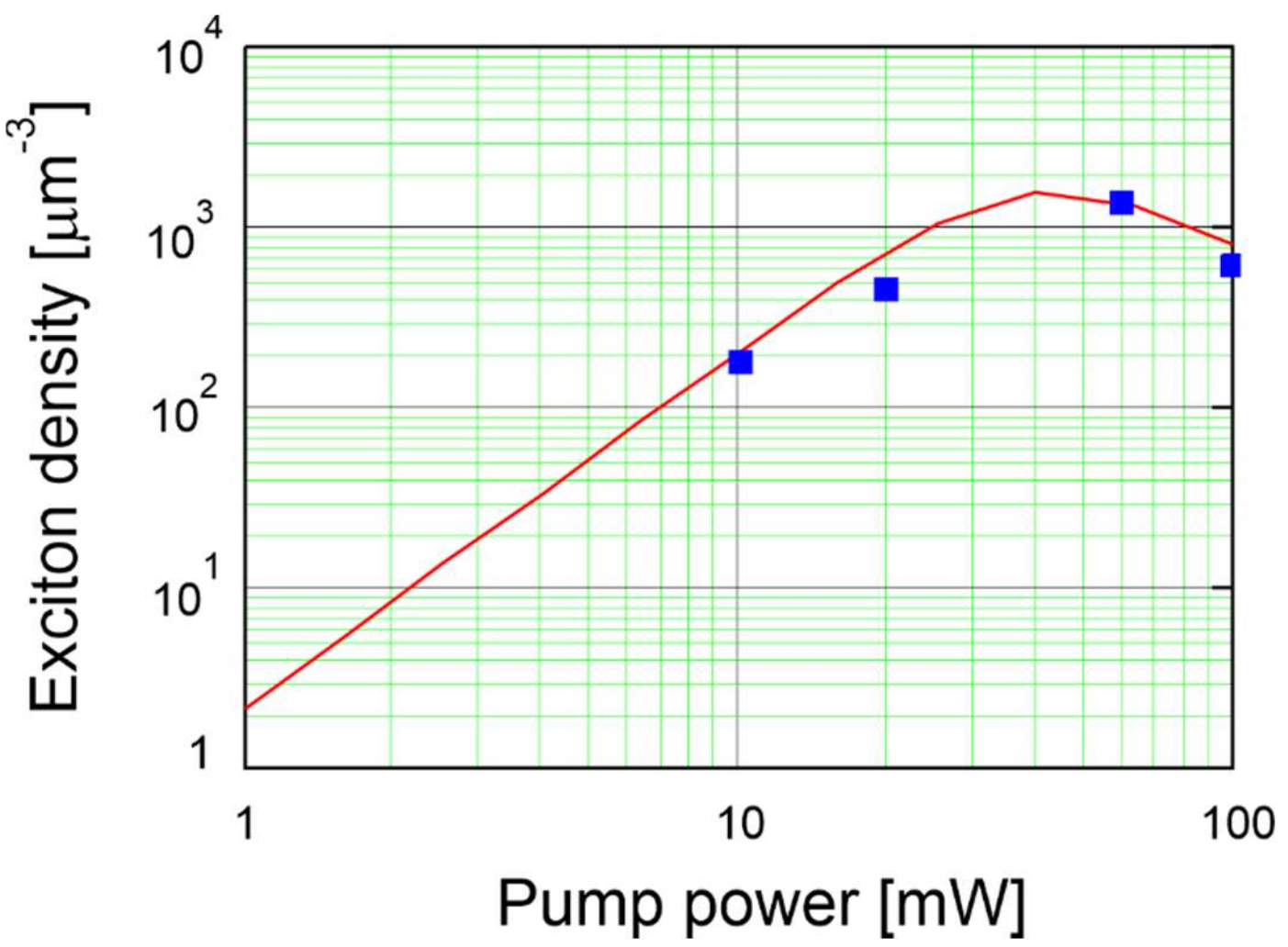

Fig. B2: Average exciton density from 2PA. The filled squares are obtained from the experiments by using Eq. (1.27) , while the full line is obtained from the quantitative analysis (Eq. (1.33)).